\documentclass[fleqn,usenatbib]{mnras}

\usepackage{newtxtext,newtxmath}

\usepackage[T1]{fontenc}

\DeclareRobustCommand{\VAN}[3]{#2}
\let\VANthebibliography\thebibliography
\def\thebibliography{\DeclareRobustCommand{\VAN}[3]{##3}\VANthebibliography}

\usepackage{graphicx}	
\usepackage{amsmath}	

\title[Unravelling structures of magnetised MCs]{Unravelling the structure of magnetised molecular clouds with SILCC-Zoom: sheets, filaments and fragmentation}

\author[S. Ganguly et al.]{
Shashwata Ganguly,$^{1}$\thanks{E-mail: ganguly@ph1.uni-koeln.de}
S. Walch,$^{1,2}$
D. Seifried,$^{1,2}$
S. D. Clarke$^{1,3}$
and M. Weis$^{1}$
\\
$^{1}$I. Physikalisches Insitut, Universit{\"a}t zu K{\"o}ln, Z\"ulpicher Str. 77, 50937 K\"oln, Germany\\
$^{2}$Cologne Centre for Data and Simulation Science, University of Cologne, Cologne, Germany\\
$^{3}$Academia Sinica, Institute of Astronomy and Astrophysics, Taipei, Taiwan\\
}

\date{Accepted XXX. Received YYY; in original form ZZZ}

\pubyear{2022}

\begin{document}
\label{firstpage}
\pagerange{\pageref{firstpage}--\pageref{lastpage}}
\maketitle

\begin{abstract}
To what extent magnetic fields affect how molecular clouds (MCs) fragment and create dense structures is an open question. We present a numerical study of cloud fragmentation using the SILCC-Zoom simulations. These simulations follow the self-consistent formation of MCs in a few hundred parsec sized region of a stratified galactic disc; and include magnetic fields, self-gravity, supernova-driven turbulence, as well as a non-equilibrium chemical network. To discern the role of magnetic fields in the evolution of MCs, we study seven simulated clouds, five with magnetic fields, and two without, with a maximum resolution of 0.1 parsec. Using a dendrogram we identify hierarchical structures which form within the clouds. Overall, the magnetised clouds have more mass in a diffuse envelope with a number density between 1-100~cm$^{-3}$. We find that six out of seven clouds are sheet-like on the largest scales, as also found in recent observations, and with filamentary structures embedded within, consistent with the bubble-driven MC formation mechanism. Hydrodynamic simulations tend to produce more sheet-like structures also on smaller scales, while the presence of magnetic fields promotes filament formation. Analysing cloud energetics, we find that magnetic fields are dynamically important for less dense, mostly but not exclusively atomic structures (typically up to $\sim 100 - 1000$~cm$^{-3}$), while the denser, potentially star-forming structures are energetically dominated by self-gravity and turbulence. In addition, we compute the magnetic surface term and demonstrate that it is generally confining, and some atomic structures are even magnetically held together. In general, magnetic fields delay the cloud evolution and fragmentation by $\sim$~1~Myr.
\end{abstract}



\section{Introduction}\label{sec:intro}
Magnetic fields are ubiquitous in the interstellar medium \citep[ISM,][]{crutcher_observations_2003, heiles_millennium_2005, fletcher_magnetic_2011, beck_magnetic_2015}. Since the discovery of interstellar magnetic fields by \citet{hiltner_polarization_1951} and \citet{hall_polarization_1951}, they have been known to be integral to the dynamical evolution of the ISM. Magnetic fields, however, are also notoriously difficult to measure accurately and model theoretically. Decades of dedicated observations have resulted in a good understanding of the morphology and strength of the magnetic field in different ISM phases \citep[][]{crutcher_magnetic_1999, bourke_oh_2001, heiles_crutcher_magnetic_2005, troland_magnetic_2008, crutcher_magnetic_2012, beck_magnetic_2015, collaboration_planck_2020, lopez-rodriguez_extragalactic_2023}. \par
However, the exact nature of how magnetic fields affect molecular cloud (MC) formation and evolution is an open question and subject of intense scrutiny \citep[see e.g. reviews by][]{crutcher_magnetic_2012,hennebelle_role_2019,girichidis_physical_2020, pattle_magnetic_2022}. Various numerical studies have performed detailed analysis on the interplay of magnetic fields with other physical processes (e.g. turbulence, thermal pressure) in order to determine how MCs are shaped, formed, and how they evolve \citep[e.g.][]{heitsch_magnetic_2001, federrath_star_2012, walch_silcc_2015, kortgen_impact_2015, girichidis_launching_2016, kortgen_origin_2018, seifried_silcc-zoom_2019, ibanez-mejia_gravity_2022}. \par
On galactic scales, ordered magnetic fields have been observed, with a correlation between the direction of the spiral arms and the magnetic field \citep[][]{beck_galactic_2009, fletcher_magnetic_2011, li_alignment_2011}. In the diffuse ISM, the magnetic field strength, $B$, does not show any correlation with the density for number densities of up to roughly 300 cm$^{-3}$ \citep[][]{crutcher_magnetic_2010}. Above these densities, \citet{crutcher_magnetic_2010} find $B\propto\rho^{\kappa}$, with $\kappa\approx 2/3$, consistent with sub-dominant magnetic field strengths, although there remains considerable scatter in the observations.\par
The lack of correlation between the strength of the magnetic field and the density of the ambient medium implies that in the diffuse ISM, magnetic fields can channelise gas flows along the field lines and therefore influence the environment in which MCs form. \citet{pardi_magnetic_2017} show that magnetic fields are more likely to cause a smoother gas distribution, while \citet{molina_density_2012} find that they are more likely to affect the dynamics of lower-density gas. Magnetic fields can add to the thermal pressure exerted by the gas and slow down the formation of dense gas \citep[][]{hill_vertical_2012}, as well as molecular gas \citep[][]{girichidis_magnetic_2018, seifried_silcc-zoom_2020}. A sufficiently strong magnetic field can prevent the collapse of a MC altogether \citep[][]{mouschovias_magnetic_1991, spitzer_physical_1978} or slow down cloud evolution \citep[][]{heitsch_magnetic_2001, padoan_star_2011, federrath_star_2012, ibanez-mejia_gravity_2022}. \par
In terms of morphology, they can facilitate the formation of elongated filamentary structures \citep[][]{hacar_initial_2022, pineda_bubbles_2022} and are essential in understanding the filamentary nature of the ISM \citep[see e.g.][]{bally_galactic_1987, andre_filamentary_2014}. The direction of such elongation relative to the direction of the magnetic field is a matter of great active research \citep[e.g.][]{soler_what_2017, seifried_parallel_2020}. In the lower density range, for sub-Alfv\'enic gas, anisotropic turbulence can lead to structures elongated parallel to field lines. In contrast, at higher densities, magnetic fields can channelise flows along field lines and therefore facilitate structures perpendicular to the field direction. \par
Magnetic fields are likely to also affect the fragmentation of clouds and cloud cores. \citet{commercon_collapse_2011} find that fragments in magnetized cloud cores are more massive compared to those formed without magnetic fields. Although the probability density function (PDF) of lower density gas is found to be different in the presence of magnetic fields \citep[][]{molina_density_2012}, the high density, potentially star-forming part does not seem to significantly affected \citep[][]{klessen_formation_2001, slyz_towards_2005, girichidis_evolution_2014, schneider_understanding_2015}. \par
In this work, we perform a numerical investigation of the role that magnetic fields play in the formation and shaping of density structures within MCs. We do a detailed analysis of realistic MC simulations based on the SILCC-Zoom simulations \citep[][]{seifried_silcc-zoom_2017} by comparing the morphological, dynamical, and fragmentation properties in seven simulated clouds, five with magnetic fields (magnetohydrodynamic or MHD clouds) and two without (hydrodynamic or HD clouds). \par 
The paper is structured as follows: In Section \ref{sec:methods}, we outline the numerical setup of the simulation. Section \ref{sec:classification} discusses the procedure for identifying and classifying structures \citep{ganguly_silcc-zoom_2022}. We highlight the differences density PDFs between HD and MHD clouds in Section \ref{sec:bulk}. The morphological properties of the obtained structures are presented in Section \ref{sec:morphology}. We find all the MCs to be sheet-like on the largest scales (tens of parsecs). On smaller scales, we see that the presence of magnetic fields enhances the formation of filamentary over sheet-like sub-structures. In Section \ref{sec:dynamics_and_fragmentation}, we analyse the dynamics and energetic balance of magnetized structures and relate them to the fragmentation of cloud sub-structures. We find that the presence of magnetic fields slows down cloud evolution and, in particular, leads to more massive fragments at low to intermediate densities (<100~cm$^{-3}$). We attempt to make an order of magnitude estimate of this slow-down effect in Section~\ref{sec:delay}. Finally, we present the summary of our findings in Section \ref{sec:conclusion}.\par

\section{Numerical methods and simulation}\label{sec:methods}
We present here results based on the SILCC-Zoom simulations \citep[][]{seifried_silcc-zoom_2017, seifried_silcc-zoom_2019}. The SILCC-Zoom simulations are MCs with realistic boundary conditions, generated by embedding the clouds within the SILCC simulations of multi-phase interstellar gas, thus having realistic initial conditions \citep[][]{walch_silcc_2015,girichidis_silcc_2016}. In this section, we highlight some key features of the simulations. More details on the simulations can be found in \citet{seifried_silcc-zoom_2017} and \citet{seifried_silcc-zoom_2019}.\par
All simulations were executed using the adaptive mesh refinement code FLASH, version 4 \citep[][]{fryxell_flash_2000, dubey_flash_2008}, which solves the ideal MHD equations for an ideal fluid. If we consider a fluid parcel of density $\rho$, velocity $\mathbf{v}$, total energy $e_{\mathrm{tot}}$, and magnetic field vector $\mathbf{B}$ (zero if pure hydrodynamics), these are given as follows:
\begin{gather}
    \frac{\partial\rho}{\partial t} + \nabla\cdot(\rho\mathbf{v})=0, \label{eq:continuity}\\
    \frac{\partial\rho\mathbf{v}}{\partial t} + \nabla\cdot \left[\rho \mathbf{v}\otimes\mathbf{v} + \left(P+\frac{B^2}{8\pi}\right)\mathbf{I} - \frac{\mathbf{B}\otimes\mathbf{B}}{4\pi}\right]=\rho\mathbf{g}, \label{eq:momentum}\\
    \frac{\partial e_{\mathrm{tot}}}{\partial t} + \nabla\cdot\left[\left(e_{\mathrm{tot}}+P\right)\mathbf{v} - \frac{(\mathbf{B}\cdot\mathbf{v})\mathbf{B}}{4\pi}\right] = \rho \mathbf{v}\cdot\mathbf{g} + \Dot{u}_{\mathrm{heat}}, \label{eq:energy}\\
    \frac{\partial\mathbf{B}}{\partial t} - \nabla \times (\mathbf{v} \times \mathbf{B}) = 0. \label{eq:magnetic}
\end{gather}
Here, Eqs.~\ref{eq:continuity} to \ref{eq:magnetic} represent the conservation of mass, momentum, energy, and magnetic flux, respectively. $P$ represents the thermal pressure, $\mathbf{g}$ is the local gravitational acceleration obtained from solving the Poisson equation, $u$ is the internal energy, and $\Dot{u}_{\mathrm{heat}}$ is the internal energy input rate due to the combination of heating and cooling processes. The $\otimes$ is the outer product (i.e. $(\mathbf{a}\otimes\mathbf{b})_{ij} = a_ib_j$). \par 
The total energy and the pressure are computed as follows:
\begin{gather}
    e_{\mathrm{tot}} = u + \frac{1}{2}\rho v^2 + \frac{1}{8\pi}B^2,\\
    P = (\gamma-1)u,
\end{gather}
with $\gamma$ being the adiabatic index. \par 
Here, we present results from runs with and without magnetic fields. The MHD simulations shown are performed using an entropy-stable solver that guarantees minimum possible dissipation \citep[][]{derigs_novel_2016, derigs_ideal_2018}. The hydrodynamic simulations have been performed using the MHD Bouchut 5-wave solver \citep{bouchut_multiwave_2007, waagan_positive_2009} that guarantees positive entropy and density. The magnetic field strength has been set to zero for these runs. \par
All simulations include self-gravity as well as an external galactic potential due to the presence of old stars. This external potential is calculated assuming a stellar population density of $\Sigma_{\rm star} = 30~\mathrm{M}_{\odot}~\mathrm{pc}^{-2}$, a sech$^2$ vertical profile and a scale height of 100~pc, according to \citet{spitzer_dynamics_1942}. The self-gravity of the gas is calculated using a tree-based algorithm \citep[][]{wunsch_tree-based_2018}.  \par
The entire simulation domain consists of a box of 500~pc~$\times$~500~pc~$\times$~$\pm$~5~kpc size, with the long axis representing the vertical $z-$direction of a galactic disc. The box is set with periodic boundary conditions in the $x-$ and $y-$ direction, and outflow boundary condition in the $z-$direction. The initial gas surface density is set to $\Sigma_{\rm gas} = 10\ \mathrm{M}_{\odot}~\mathrm{pc}^{-2}$ which corresponds to solar neighbourhood conditions. The vertical distribution of the gas is modelled as a Gaussian, i.e. $\rho = \rho_0\ \mathrm{exp}(-z^2/2h_{z}^2)$, where $h_z$=30 pc is the scale height and $\rho_0 = 9 \times 10^{-24}$~g~cm$^{-3}$. The initial gas temperature is set to 4500 K. For runs with magnetic fields, the magnetic field is initialized along the $x-$direction, i.e. $\mathbf{B} = (B_x,0,0)$ with $B_x = B_{x,0}\sqrt{\rho (z)/\rho_0}$ and the magnetic field strength at the midplane $B_{x,0} = 3\ \mu$G. The field strength is chosen to be in accordance with recent observations \citep[e.g.][]{beck_magnetic_2013}.   \par
The turbulence in the simulations is generated by supernova explosions. The explosion rate is set to 15 SNe Myr$^{-1}$, which is consistent with the Kennicutt-Schmidt relation, which observationally determines the star formation rate surface density for a given gas surface density \citep[][]{schmidt_rate_1959, kennicutt_global_1998}. 50\% of the supernovae are placed following a Gaussian random distribution along the $z-$direction up to a height of 50 pc, while the other 50\% are placed at density peaks of the gas. This prescription of supernova driving creates a multi-phase turbulent ISM which can be used as initial conditions for the zoom-in simulations \citep[][]{walch_silcc_2015,girichidis_silcc_2016}. \par
Apart from the dynamics of the gas, we also model its chemical evolution using a simplified non-equilibrium chemical network based on hydrogen and carbon chemistry \citep[][]{nelson_dynamics_1997, glover_simulating_2007, glover_modelling_2010}. For this purpose, we follow the abundance of H$^+$, H, H$_2$, CO, C$^+$, e$^-$, and O. At the beginning of the simulation, all hydrogen in the disc midplane is neutral and carbon is in its ionized form (i.e. H and C$^+$, respectively).  \par
To correctly model the chemistry of the gas, we include an interstellar radiation field (ISRF) of strength $G_0=1.7$ in Habing units \citep[][]{habing_interstellar_1968, draine_photoelectric_1978}. The attenuation of this radiation field is taken into consideration by computing the true optical depth inside any given point in the simulation domain. This is computed as follows:
\begin{equation}\label{eq:av_3d}
    \mathrm{A_{V,3D} = -\frac{1}{2.5}ln \left[\frac{1}{{\it N}_{PIX}} \sum_{i=1}^{{\it N}_{PIX}} exp \left(-2.5\frac{N_{H,tot,i}}{1.87 \times 10^{21}\ \mathrm{cm}^{-2}}\right)\right]},
\end{equation}
where the sum is carried over each \textsc{Healpix} pixel, with $N_{\rm PIX}$ being the total number of such pixels (here 48), and $N_{\mathrm{H,tot},i}$ is the column density computed for the $i-$th pixel. In essence, for any given point, we compute the column density along various lines of sight and use that for an effective $\mathrm{A_{V,3D}}$. The averaging is performed in an exponential manner because the intensity of radiation decreases in an exponential manner due to extinction caused by the gas column density along the line of sight. The calculation for this is performed by the \textsc{TreeRay} \textsc{Optical Depth} module developed by \citet{wunsch_tree-based_2018}. \par
To study the formation of MCs, all supernova explosions are stopped at a certain time $t_0$. Up to this point, the maximum grid resolution is 3.9 pc. At time $t_0$, different regions are identified for the zoom-in process, primarily by determining which regions form molecular gas when the simulations are run further at the original SILCC resolution of 3.9~pc. The time $t=t_0$ refers to the start of the evolution of the different clouds and is set as an evolutionary time $t_{\rm evol}=0$. The total simulation time $t$ is related to the evolution time as 
\begin{equation}
    t = t_0 + t_{\rm evol}.
\end{equation}
From $t_{\rm evol}=0$ on, in the selected regions, the AMR grid is allowed to be refined to a higher resolution to capture structures that form as MCs. These regions are called zoom-in regions and are of primary importance to us as sites of MCs. Each SILCC simulation we run contains two such "zoom-in" boxes simultaneously. All runs present here have a maximum resolution of 0.125 pc. For details of how the zoom-in process is achieved, see \citet{seifried_silcc-zoom_2017}.\par


\section{Classification of structures} \label{sec:classification}

For the analysis presented in this work, we look at eight different cubic boxes of 62.5 pc in size, each from a different SILCC zoom-in region. These boxes are chosen by visual inspection, in order to capture the most interesting features contained in each zoom-in region. For the purpose of this work, we will refer to these cubic regions as MCs. They are named MC1-HD and MC2-HD for the two hydrodynamic clouds, and MC$x$-MHD for the MHD clouds, where $x$ is between one and six. We present some basic details of the different MCs in Table \ref{tab:cloud_info}. A projected view of all the different MCs is added in the Appendix \ref{sec:app_a}. For more information on the presented clouds, we refer the reader to \citet{seifried_silcc-zoom_2017} for the HD clouds and \citet{seifried_silcc-zoom_2019} for the MHD clouds. \par 
\begin{table}
\centering
    \begin{tabular}{l|c|c|c|c|c}
        \hline
         Run name & MHD & $t_0$  & Total mass & H$_2$ mass & $\langle \mathrm{B} \rangle$\\
         &   & [Myr] &$[10^4\,{\rm M}_{\odot}]$& $[10^4\,{\rm M}_{\odot}]$& [$\mu$G]\\
         \hline
         MC1-HD&  no& 12& 7.3 & 2.1&0 \\
         MC2-HD&  no& 12& 5.4 & 1.6&0 \\
         \hline
         MC1-MHD& yes& 16& 7.8 & 1.3&4.8 \\
         MC2-MHD& yes& 16&  6.2 & 0.86&3.9 \\
         (MC3-MHD$^{\rm a}$& yes& 16& 2.0 & 0.19&2.0) \\
         MC4-MHD& yes& 11.5&  6.8 & 1.2 & 6.4 \\
         MC5-MHD& yes& 11.5&  10.1 & 1.6 & 6.8\\
         MC6-MHD& yes& 16& 6.6 & 1.4 & 4.3 \\
         \hline
    \end{tabular}
    \caption{Basic information on the eight analysed simulations. From left to right we list the run name, whether magnetic fields are present or not, the time when the AMR "zoom-in" starts, as well as the total mass, molecular hydrogen mass and the average magnetic field strength at $t_{\rm evol}=2$ Myr.\newline 
    $^{\rm a}$We discard MC3-MHD from our further analysis because of its low molecular gas content and lack of interesting density features (see also Fig.~\ref{fig:projection_plot}).}
    \label{tab:cloud_info}
\end{table}
We perform a detailed analysis of the different clouds, following their evolution from $t_{\rm evol}=2$~Myr to $t_{\rm evol}=3.5$~Myr, primarily focusing on the latter time. The beginning and the end time are chosen to look at relatively early stages of structure formation in the MCs. We do not look at times earlier than 2~Myr primarily because the clouds undergo the refinement process and are not fully resolved until $t_{\rm evol}\sim 1.5$~Myr. 

\subsection{Structure identification}\label{sec:methods_structure_identification}
To identify structures in our MCs, we use a dendrogram algorithm \citep[][]{rosolowsky_structural_2008}. Dendrogram is a model-independent method to determine hierarchical structures in two and three dimensions. 
Since we are interested in 3-dimensional structures, we perform the dendrogram analysis on 3-dimensional density cubes. We do not use the 3D AMR grid structure inherent in the data, but rather convert it into a uniform mesh at 0.125 and 0.25 pc resolution (see also Table~\ref{tab:dendrogram_type}). \par

Given an initial density field, $\rho$, the dendrogram essentially depends on three free parameters: the initial starting threshold, $\rho_0$, the density jump, $\Delta \rho$, and the minimum number of cells that need to be included in any structure, $N_{\rm cells}$. Due to high density contrasts, we build the dendrogram tree on the logarithmic density profile of the gas, and therefore have used density bins of $\Delta \mathrm{log}_{10}\ \rho$, rather than $\Delta \rho$. In addition to the three parameters mentioned, we can choose a pruning peak, $\rho_{\rm prune}$, to allow the dendrogram to create new structures only when such a structure will have peak density $\rho_{\rm peak}>\rho_{\rm prune}$, although this has not been used in the present work. Using these parameters, the dendrogram algorithm allows us to define volumes of gas as structures in a hierarchical tree, primarily defined by their threshold density $\rho_{\rm thr}$, which is the minimum density value inside a given structure. This can be thought of as equivalent to contour values for two dimensional maps. The hierarchy is characterised by different dendrogram \textit{branches}, where a branch is a given dendrogram structure and all its parent structures, up to the largest and most diffuse ancestor in the dendrogram tree.\par 
For probing both the higher and lower density ends of the data, we perform two dendrgram analyses on the same regions: a higher density dendrogram analysis performed at a resolution of 0.125 pc for probing gas above densities of 10$^{-22}$ g cm$^{-3}$ (referred to as \textit{high-den}), and a lower density analysis performed at 0.25 pc for gas between the densities of 10$^{-24}$ and 10$^{-22}$ g cm$^{-3}$ (referred to as \textit{low-den}). The \textit{low-den} values are computed as volume averaged values from the higher resolution grid. We present the dendrogram parameters used for both analyses in Table \ref{tab:dendrogram_type}. \par
\begin{table*}
    \centering
    \begin{tabular}{ccccccc}
    \hline
         dendrogram&Resolution&$\rho_0$&$\Delta$ log$_{10}\ \rho$&$N_{\rm cells}$&$\rho_{\rm prune}$&additional \\
         type&[pc]&[g cm$^{-3}$]&&&[g cm$^{-3}$]&criteria\\
         \hline
         \textit{high-den}&0.125&$10^{-22}$&0.1&100&None&None\\         
         \textit{low-den}&0.25&$10^{-24}$&0.2&100&None&$\rho_{\rm thr}<10^{-22}$ g cm$^{-3}$\\
  \hline   
    \end{tabular}
    \caption{Information on the parameters used for the two different kinds of dendrogram analyses. From left to right are: the type of dendrogram, the grid resolution at which it is performed, the starting density, the logarithmic density jump, the minimum number of cells in structures, the density of the pruning peak used, and if any additional criteria were used to select structures.}
    \label{tab:dendrogram_type}
\end{table*}
In addition to the difference in the basic parameters between the two dendrogram analyses, we remove all structures with $\rho_{\rm thr}>10^{-22}$~g~cm$^{-3}$ for the \textit{low-den} analysis. This is done in order to avoid double counting of structures.\par
The parameter values mentioned in Table \ref{tab:dendrogram_type} have been chosen from a mixture of practical considerations, such as CPU memory, computation time, and through trial and error. We note that in principle the same analysis could be performed by a single dendrogram analysis at $\rho_{\rm thr}=10^{-24}$~g~cm$^{-3}$ at the highest resolution of 0.125 pc. However, the computation cost of such an analysis was prohibitive in our case. Combining the \textit{high-den} and \textit{low-den} dendrogram analyses allows us to probe a much higher density range than would be otherwise possible. \par 
In terms of the parameters used, we have seen no unexpected change in the results by changing the free parameters within a reasonable range. We refer the reader to our companion paper \citep[][]{ganguly_silcc-zoom_2022} for a more thorough discussion of the effect of altering the parameter values on the analysis. Overall, we find that changing the parameters, while resulting in a varying number of obtained structures, leaves the statistical properties of the structures virtually unaffected. \par
\begin{figure*}
    \centering
    \includegraphics[width=0.99\textwidth]{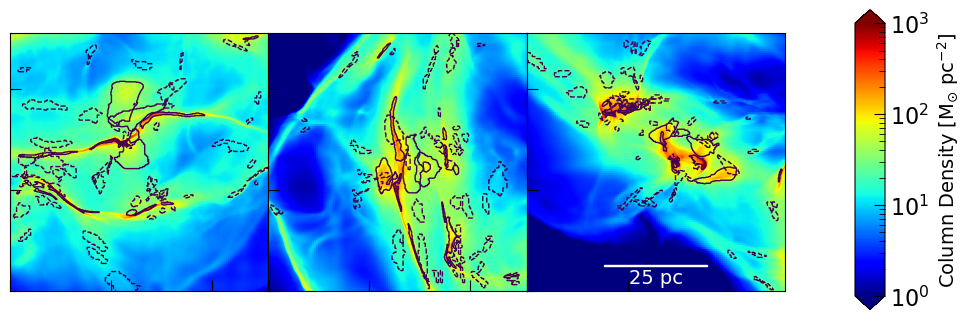}
    \caption{Left to right: Projections of MC1-MHD at $t_{\rm evol}=$3.5 Myr along the $x$-, $y$-, and $z$-axis, respectively. The contours show the projections of the leaf dendrogram structures along the same axis. Molecular structures ($> 50\% \; \mathrm{H}_2$ mass fraction) are plotted with solid, and atomic structures ($< 50\% \; \mathrm{H}_2$ mass fraction) are plotted with dashed lines. The molecular structures nicely trace the dense spine of the two main filaments, while the atomic structures mostly represent the envelope.}
    \label{fig:contours}
\end{figure*}

An example of the leaf density structures (structures that contain no  further sub-structures) from the dendrogram analysis can be seen in Fig. \ref{fig:contours} for MC1-MHD at $t_{\rm evol}=3.5$~Myr, as contours over column density maps. The three panels show, from left to right, the cloud projected along the $x-$, $y-$ and $z-$direction. The contours are drawn as projections of the 3D dendrogram structure outlines in the projected direction. We distinguish between structures depending on their molecular H$_2$ content, by plotting structures with over 50\% of their total hydrogen mass in molecular form (referred to as molecular structures) in solid lines and otherwise in dashed lines (referred to as atomic structures). \par
Due to the nature of the dendrogram algorithm, there are some structures which touch the edge of the box. This can lead to structures whose morphology is determined by their proximity to the edge. To avoid this, we do not classify the morphology of any structures which have more than 5\% of their surface cells touching any edge. This is relevant especially for the large-scale structures from the \textit{low-den} dendrogram analysis. However, they can still be of interest while considering cloud dynamics, and in such a case we add them as an additional category of "unclassified". While in a different context, \citet{alves_shapes_2017} have shown the importance of having closed contours while studying 2D maps. We have attempted to follow the same principle here as much as possible.

\subsection{Structure classification}\label{sec:methods_structure_classification}
Once we obtain the tree of dendrogram density sub-structures, we aim to classify their morphology. For each structure, we compute an equivalent ellipsoid that has the same mass and the same moments of inertia (MOI) as the original structure. We then use the axes lengths of this equivalent ellipsoid to classify the shape of the different structures. \par
Let us consider a uniform density ellipsoid of mass $M$ and semi-axes lengths $a,\ b,\ c$ with $a\geq b\geq c$. The moments of inertia along the three principal axes will be given as follows:
\begin{equation}
    \begin{split}
        I_a &= \frac{1}{5} M(b^2+c^2),\\
        I_b &= \frac{1}{5} M(c^2+a^2),\\
        I_c &= \frac{1}{5} M(a^2+b^2),
    \end{split}
\end{equation}
where $I_c\geq I_b\geq I_a$. If we now compute the principal moments of inertia of our given dendrogram structure to be $A$, $B$ and $C$, respectively, then the ellipsoid has an equivalent moment of inertia if
\begin{equation}
        A =I_a,\ B=I_b,\ C=I_c.
\end{equation}
This leads to the following equation for computing the axis lengths of the equivalent ellipsoids:
\begin{equation}\label{eq:fit_ellipsoid}
    \begin{split}
        a &= \sqrt{\frac{5}{2M}(B+C-A)}\ ,\\
        b &= \sqrt{\frac{5}{2M}(C+A-B)}\ ,\\
        c &= \sqrt{\frac{5}{2M}(A+B-C)}\ .
    \end{split}
\end{equation}
We then use the aspect ratio of the semi-axes of the corresponding ellipsoid and the position of the center of mass (COM) of the structure relative to its boundary (i.e. whether the COM is contained by the structure itself) to categorise the different structures into four categories - sheets, curved sheets (referred to as sheet\_c in this paper), filaments, and spheroids: \par
\begin{equation}
    \begin{split}
        &\textbf{sheet: } \frac{a}{b} \leq f_{\rm asp}, \frac{a}{c} > f_{\rm asp}\\
        &\textbf{filament: } \frac{a}{b} > f_{\rm asp}\\
        &\textbf{spheroidal: } \frac{a}{c} \leq f_{\rm asp}, \text{ contains its own COM} \\
        &\textbf{sheet\_c: }\frac{a}{c} \leq f_{\rm asp}, \text{ does not contain its own COM}
        \end{split}
\end{equation}
where we set the aspect ratio factor $f_{\rm asp}=3$.\par
The inclusion of the COM criterion in addition to the ratio of the ellipsoid axes help us deal with especially the larger-scale structures which can be highly curved. A highly curved sheet could have comparable MOI eigenvalues along the different eigen-directions, but would not contain its own COM. We highlight some visual examples of such highly curved sheet-like structures when we discuss the large scale morphology of our clouds in Section~\ref{sec:morphology}. In contrast to curved sheets, a spheroidal structure would contain its own COM. \par
Apart from using the normal moment of inertia, we also perform the classification by computing a volume-weighted moment of inertia, where we compute the moment of inertia of the structures (the quantities $A$, $B$ and $C$) by assuming the structure is of the same mass but with uniform density, but find statistically little to no difference in the resulting morphologies. \par
The discussion above highlights some possible caveats of our method. If we have a situation of multiple crossing filaments (hub-like structure), or parallel filaments joined by a more diffuse intermediate medium - the method will identify it as a sheet-like structure splitting into filaments in the dendrogram tree hierarchy. We must therefore emphasise that our definition of a sheet in this context is more general and contains also situations where multiple filamentary structures are connected by a more diffuse medium. Further, for highly curved structures, it is possible that the simple fit ellipsoid method may not result in a good description of the ellipsoid axis lengths. 

\section{Density distribution and magnetic fields}\label{sec:bulk}
We first consider the bulk properties of the different MCs to quantify the differences between the hydrodynamic and MHD clouds. From Table \ref{tab:cloud_info}, we see that the volume-weighted root-mean-square average magnetic field strength for all MHD clouds is comparable and varies between 3.9-6.8~$\mu$G. These values are slightly higher than the initial magnetic field strength $B_{x,0} =3\,\mu{\rm G}$. The cloud masses and their H$_2$ masses are also within a factor of roughly 2 to each other (with the exception of MC3-MHD, see below). For a view of the time evolution of the total and H$_2$ masses, as well as the H$_2$ mass fraction, we refer the reader to Appendix~\ref{sec:app_a}. \par 
MC3-MHD stands out as it has a much lower H$_2$ mass and H$_2$ mass fraction compared to the other clouds (Table~\ref{tab:cloud_info}). Visual inspection of this cloud shows that its structures are still diffuse and not as prominent, suggesting that it perhaps needs much longer to collapse, or may not collapse at all (see Fig.~\ref{fig:projection_plot}, bottom row left). Its molecular content remains at a roughly constant level of 10\% throughout. Since we are interested primarily in the problem of density structures that eventually form stars, we exclude MC3-MHD from further analysis considering its unevolved state and low molecular content. \par

\begin{figure}
    \centering
    \includegraphics[width=0.49\textwidth]{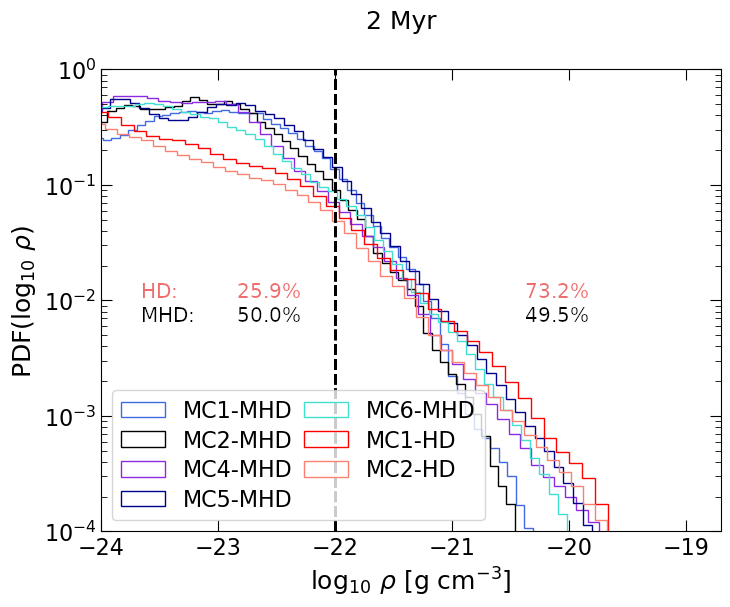}
    \includegraphics[width=0.49\textwidth]{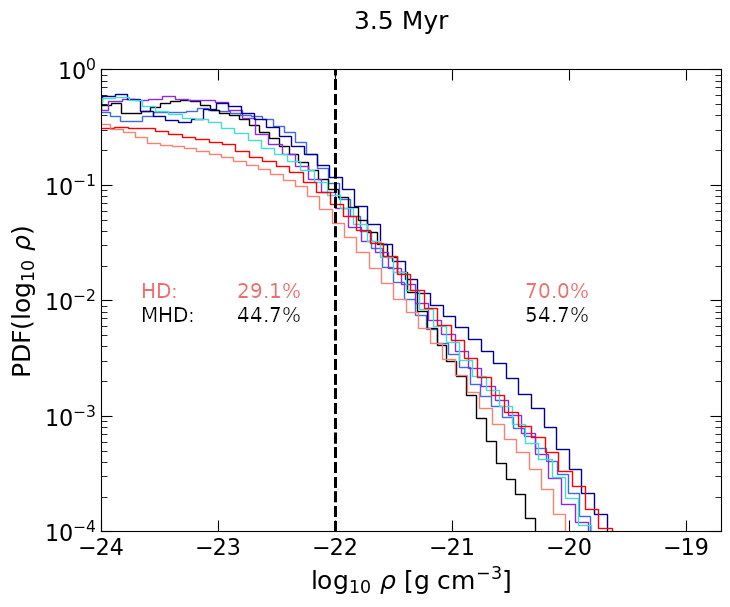}
    \caption{Volume-weighted density PDF for different HD and MHD clouds $t_{\rm evol}=2$~Myr (top) and 3.5~Myr (bottom). The density range shown is used for a dendrogram analysis, and contains more than 99\% of the total mass of the clouds. The two hydrodynamic clouds are plotted in reddish lines. The vertical line demarcate the boundaries of the \textit{high-den} ($>10^{-22}$~g~cm$^{-3}$) and the \textit{low-den} (between $10^{-24}-10^{-22}$~g~cm$^{-3}$) dendrogram analyses (see also Table \ref{tab:dendrogram_type}). The MHD clouds have more fraction of gas in the density range between roughly $10^{-24}$ and $10^{-22}$ g cm$^{-3}$, or between approximately 1 and 100 cm$^{-3}$.}
    \label{fig:density_pdf}
\end{figure}

It is of interest to examine whether the mass distribution in different clouds is affected by the presence of magnetic fields. This can be seen in Fig.~\ref{fig:density_pdf}, which shows the volume-weighted density PDF of all different clouds at $t_{\rm evol}=2$~Myr (top) and $t_{\rm evol}=3.5$~Myr (bottom) in the density range probed by the dendrogram analysis ($>10^{-24}$~g~cm$^{-3}$). The respective density PDFs for the full density range can be found in Appendix~\ref{sec:app_pdf}. The two hydrodynamic clouds are plotted using reddish lines (red and salmon), while the magnetised clouds are shown using darker colours. For all clouds, the shown density range contains more than 99\% of their total mass. \par
From Fig.~\ref{fig:density_pdf}, we see that between $10^{-24}$ and $10^{-22}$ $\mathrm{g\ cm^{-3}}$, corresponding to the rough number densities between 1 and 100~$\mathrm{cm^{-3}}$, the MHD clouds contain much more gas. This is more prominent at $t_{\rm evol}=2$~Myr, but remains also clearly visible at $t_{\rm evol}=3.5$~Myr. This effect can also be visually seen in the column density plots of Fig. \ref{fig:projection_plot}, where the denser parts of the hydrodynamic clouds seem to be embedded in a more rarefied medium compared to their MHD counterparts. We calculate the mass percentage at 2~Myr in different density regimes in Table~\ref{tab:mass_in_density_regimes}, which shows that, at this time, the MHD clouds contain almost 50\% of their mass between $10^{-24}$ and $10^{-22}$~$\mathrm{g\ cm^{-3}}$, in contrast to only around 26\% for the hydrodynamic MCs.\par
\begin{table}
    \centering
    \begin{tabular}{@{\extracolsep{\fill}}*{4}{c}}
    \hline
        Cloud & \multicolumn{3}{c}{Mass percentage at 2 Myr}\\
        \cline{2-4}
        \rule{0pt}{3ex}
        sample &$\frac{\rho}{{\rm g}\,{\rm cm}^{-3}}<10^{-24}$ &$10^{-24} \leq \frac{\rho}{{\rm g}\,{\rm cm}^{-3}} < 10^{-22}$&$\frac{\rho}{{\rm g}\,{\rm cm}^{-3}} \geq 10^{-22}$\\
        \hline 
        HD & 0.9 & 25.9 & 73.2 \\
        MHD & 0.5 & 50.0 & 49.5 \\
    \hline
    \end{tabular}
    \caption{The average mass percentage in different density regimes for the HD and MHD clouds at $t_{\rm evol}=2$~Myr. The MHD clouds have twice the amount of mass in the intermediate density range between $10^{-24}$~g~cm$^{-3}$ and $10^{-22}$~g~cm$^{-3}$ compared to their HD counterparts.}
    \label{tab:mass_in_density_regimes}
\end{table}
Magnetic fields in our simulations therefore play an important role in shaping the environment inside which denser, molecular, and potentially star-forming structures live. This is consistent with the picture that magnetic fields have a noticeable effect on the dynamics of low (here, $\lesssim 10^{-22}$~g~cm$^{-3}$) density gas \citep[][]{molina_density_2012}. Similar conclusions have been reached by \citet{seifried_parallel_2020} using the technique of relative orientation of magnetic fields with respect to filaments who find that this change in the relative impact of magnetic fields occurs around $\sim 100$~cm$^{-3}$. We explore the effects of magnetic fields in more detail by looking at the clouds' fragmentation properties in Section \ref{sec:fragmentation}. \par

For $\rho>10^{-22}$~g~cm$^{-3}$ we see no clear trend in the slope of the density PDF between the hydrodynamic and MHD clouds. This is consistent with simulations and observations showing that column density PDFs are not sensitive to the presence of a magnetic fields in the high column density regime \citep[][]{klessen_formation_2001, slyz_towards_2005, girichidis_evolution_2014, schneider_understanding_2015}. 

However, at $t_{\rm evol}=2$~Myr, the two hydrodynamic clouds seem to have a bit more dense gas mass (see also Table~\ref{tab:mass_in_density_regimes}), although the effect is visually far from clear. If there were a "delay" in the formation of denser gas when magnetic fields are present, this would be extremely relevant for the formation of well-shielded, molecular gas. In Table~\ref{tab:av_dist}, we show the mass above an $\mathrm{A_{V,3D}}$ (Eq.~\ref{eq:av_3d}) of 1 and 10 for one magnetised and one non-magnetised cloud of comparable mass (MC1-MHD and MC1-HD, see Table~\ref{tab:cloud_info}). Additionally, the mass-weighted PDF of $\mathrm{A_{V,3D}}$ for these two clouds is shown in Appendix~\ref{sec:app_av}, Fig.~\ref{fig:av_pdf}. From Table~\ref{tab:av_dist}, as well as from Fig.~\ref{fig:av_pdf}, we find that the amount of gas above $\mathrm{A_{V,3D}}>1$ and $\mathrm{A_{V,3D}}>10$ in MC1-MHD is consistently lower compared to MC1-HD. In Section~\ref{sec:delay}, we attempt to quantify such a delay timescale due to magnetic fields. For a more detailed analysis on the connection between magnetic fields and $\mathrm{A_{V,3D}}$, we refer the reader to \citet{seifried_silcc-zoom_2020}.\par
\begin{table}
    \centering
    \begin{tabular}{ccc}
    \hline
         cloud, time &mass above&mass above  \\
         & $\mathrm{A_{V,3D}}>1$ [\%] & $\mathrm{A_{V,3D}}>10$ [\%]\\
         \hline
         MC1-HD, 2 Myr & 41.1& 2.3\\
         MC1-HD, 3.5 Myr &44.0 & 9.8\\
         \hline
         MC1-MHD, 2 Myr & 26.6& 0\\
         MC1-MHD, 3.5 Myr & 31.7&0.8\\
         \hline
    \end{tabular}
    \caption{The percentage of mass above values of $\mathrm{A_{V,3D}}=1,\ 10$ for two similar mass clouds, MC1-HD and MC1-MHD.}
    \label{tab:av_dist}
\end{table}

\section{Morphology}\label{sec:morphology}
\begin{figure*}
    \centering
    \includegraphics[trim=250 20 180 5, clip, width=0.33\textwidth]{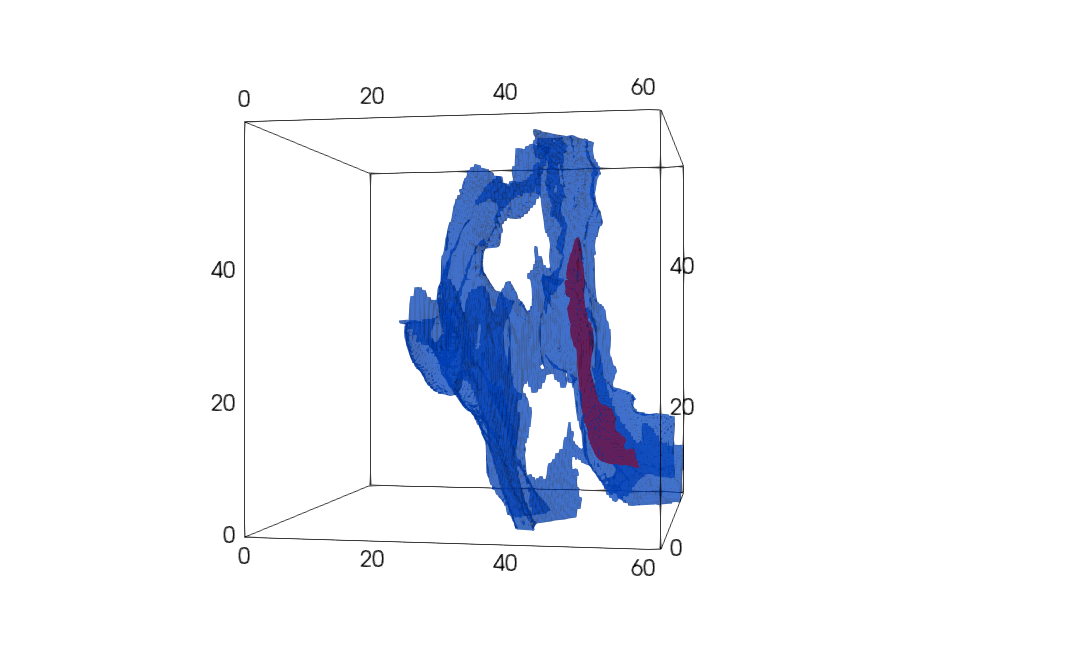}
    \includegraphics[trim=250 20 180 5, clip,width=0.33\textwidth]{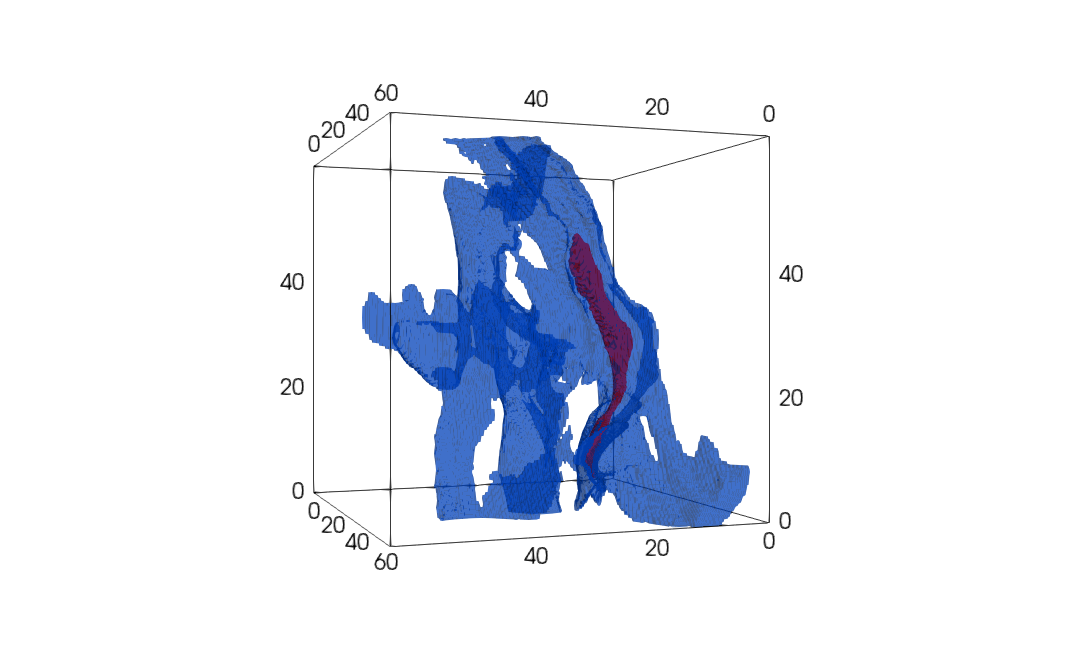}
    \includegraphics[trim=250 20 180 5, clip,width=0.33\textwidth]{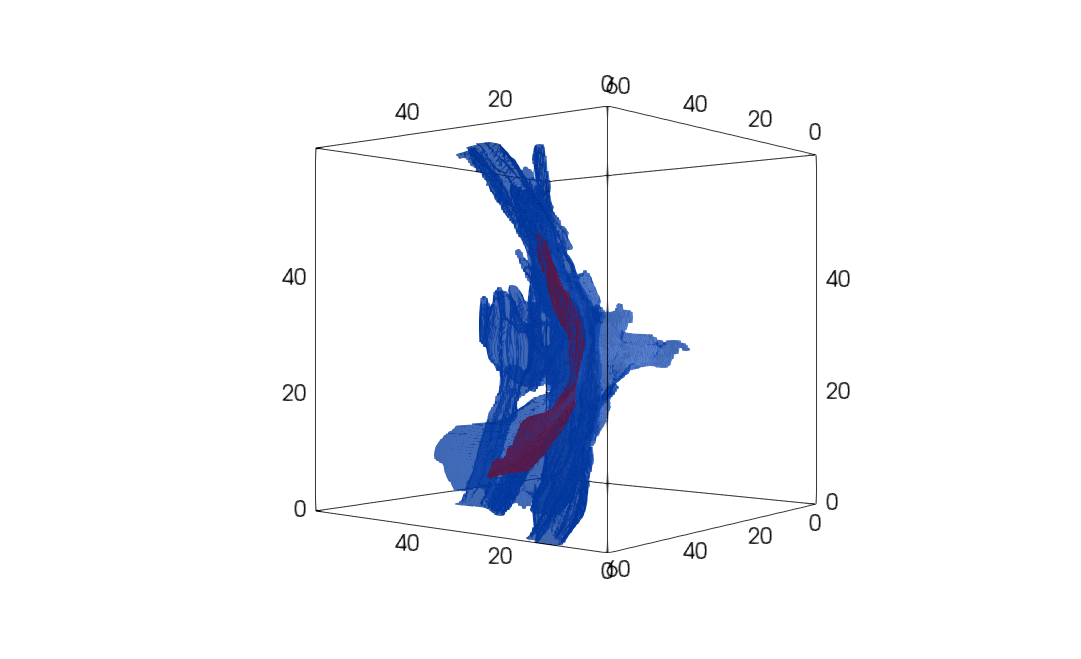}
    
    \includegraphics[trim=250 20 180 5, clip, width=0.33\textwidth]{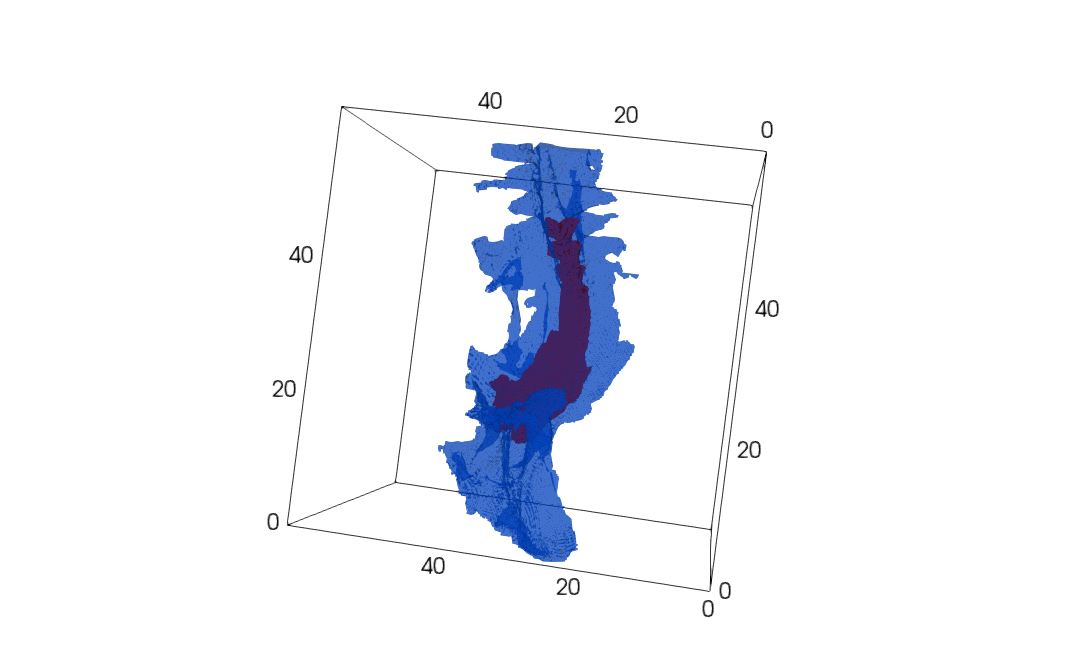}
    \includegraphics[trim=250 20 180 5, clip,width=0.33\textwidth]{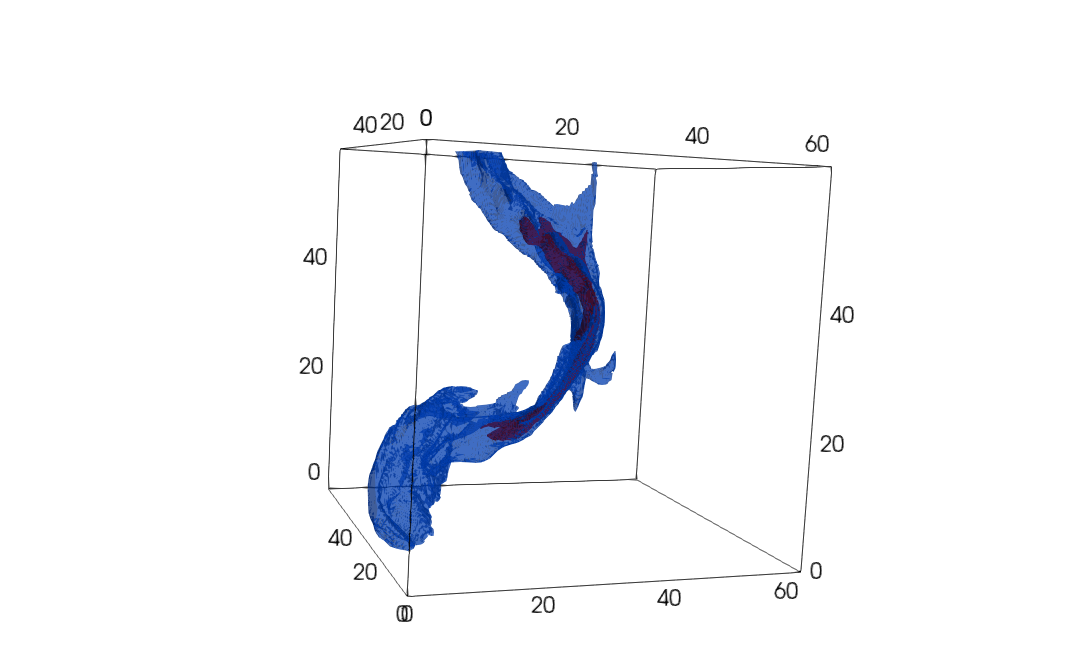}
    \includegraphics[trim=250 20 180 5, clip,width=0.33\textwidth]{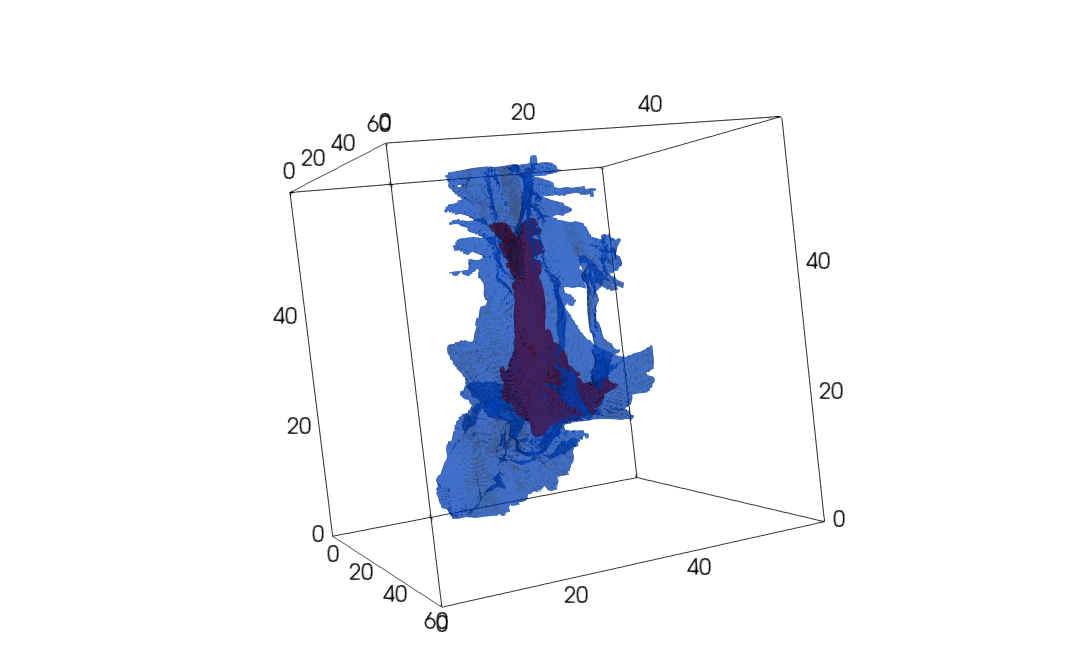}

    \includegraphics[trim=250 20 180 5, clip, width=0.33\textwidth]{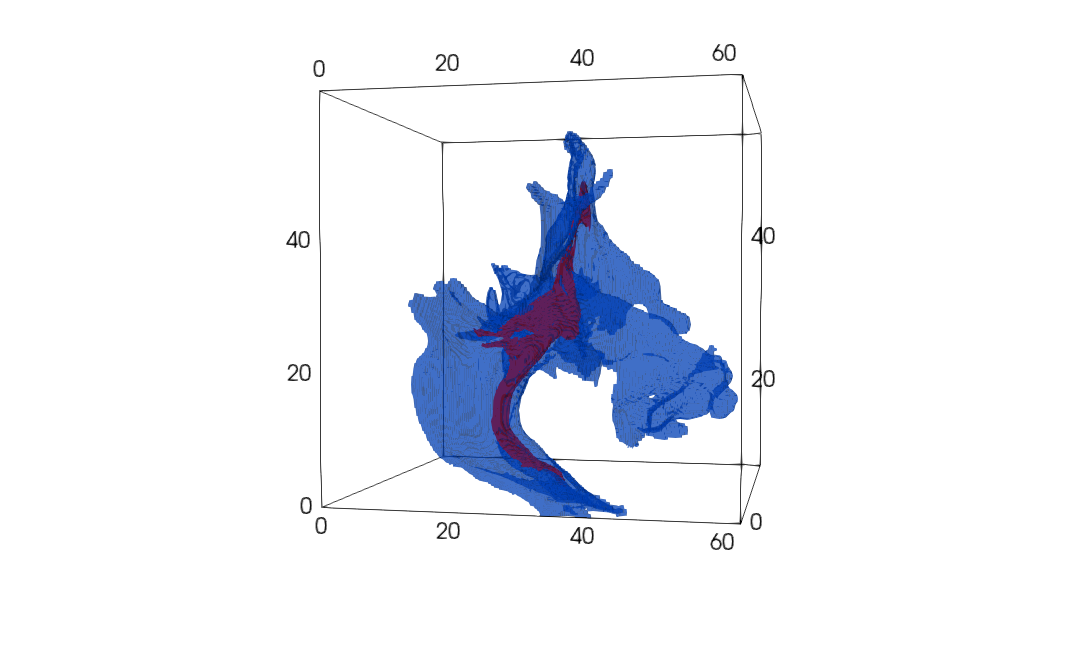}
    \includegraphics[trim=250 20 180 5, clip,width=0.33\textwidth]{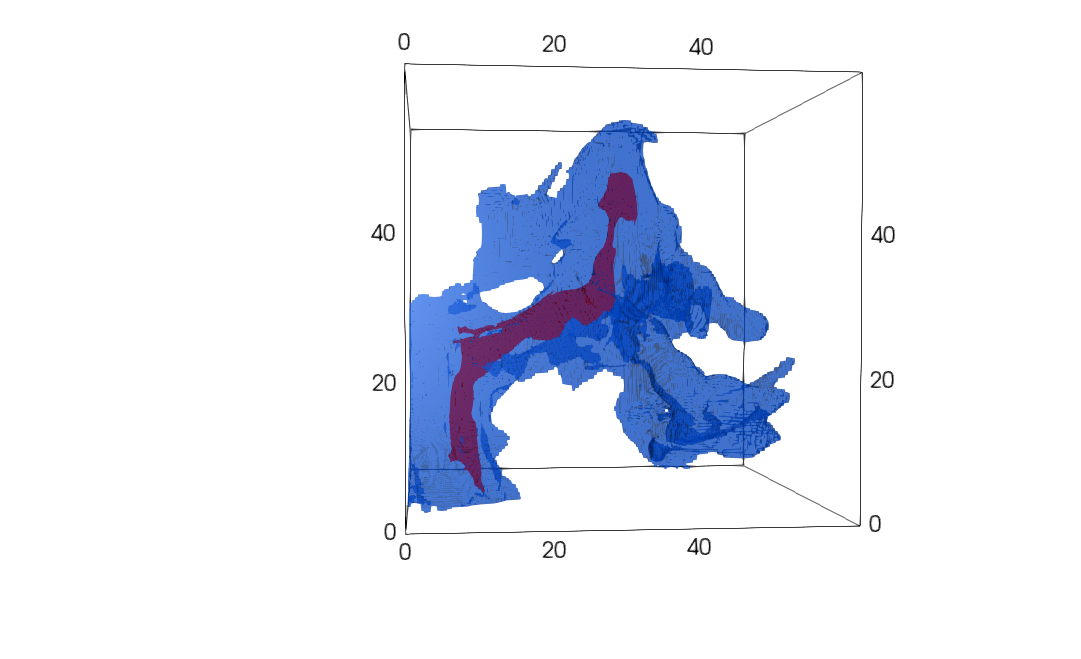}
    \includegraphics[trim=250 20 180 5, clip,width=0.33\textwidth]{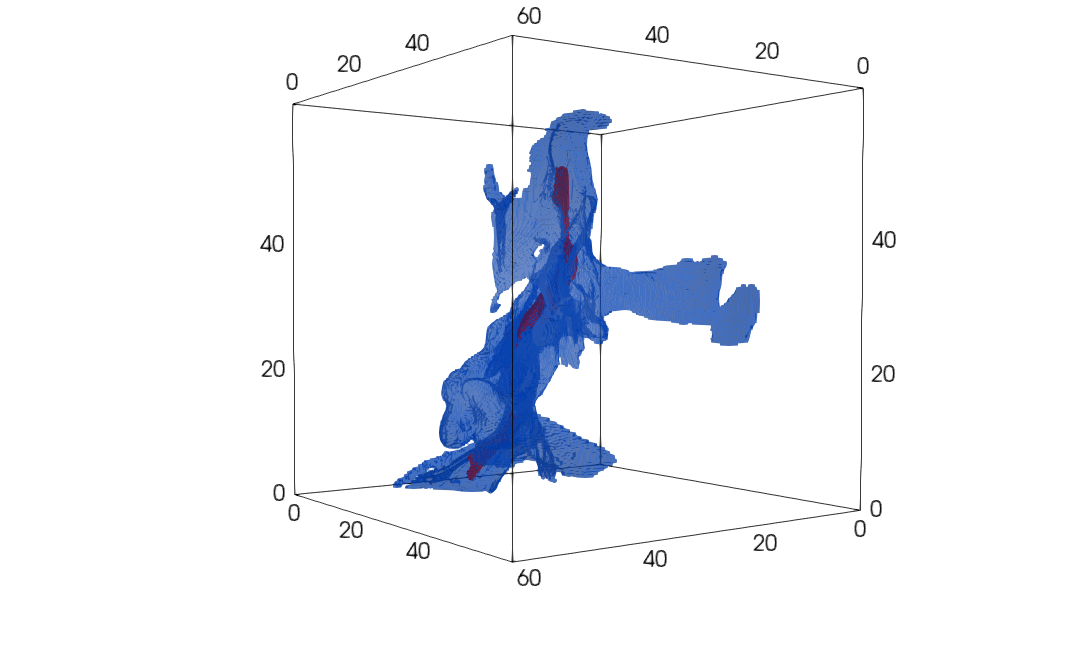}

    \caption{3D surface rendering of example large-scale dendrogram structures from the \textit{high-den} dendrogram analysis for MC1-MHD (top row), MC5-MHD (middle row), and MC6-MHD (bottom row), from different viewing angles (left to right). The blue structures represent the large-scale sheets or curved sheets at $\rho_{\rm thr} \approx 10^{-22}$~g~cm$^{-3}$, while the embedded red structures show one of the more prominent embedded filaments ($\rho_{\rm thr}$ between $10^{-20}-10^{-21}$~g~cm$^{-3}$). The units in the axes are in parsec. A video link for the various structures can be found in \url{https://hera.ph1.uni-koeln.de/~ganguly/silcc_zoom/morphology_3d/}.}
    \label{fig:3d_projection}
\end{figure*}

We perform a morphological classification of all simulated cloud structures using the method described in Section~\ref{sec:methods_structure_classification}. As an intuitive visual aid, we first present 3D surfaces of three large-scale cloud dendrogram structures\footnote{We show the largest structures from the \textit{high-den} dendrogram analysis ($\rho > 10^{-22}$~g~cm$^{-3}$) as they are on the maximum resolution and therefore capture the finer complexities of the cloud better. The large scale structures for the \textit{low-den} dendrogram analysis follow the same trend.} 
(from top to bottom: MC1-MHD, MC5-MHD, and MC6-MHD) seen from three different viewing angles (different columns) in Fig.~\ref{fig:3d_projection}. The lighter blue colour shows the large-scale structure (identified at $\rho_{\rm thr}\approx10^{-22}$~g~cm$^{-3}$) and in red, we show one of the primary embedded filamentary structures (identified using values of $\rho_{\rm thr}$ between $10^{-20} - 10^{-21}$~g~cm$^{-3}$). 
Visual inspection seems to suggest that the large-scale, lighter blue structures are rather thin and sheet-like, and indeed all three clouds shown in Fig.~\ref{fig:3d_projection} are identified as sheets or curved sheets according to the classification algorithm of Section~\ref{sec:methods_structure_classification}. This is even clearer in a video view, which can be found here\footnote{\url{https://hera.ph1.uni-koeln.de/~ganguly/silcc_zoom/}}. The visual suggestion of the clouds being sheet-like on the largest scales is also confirmed for all clouds in a quantitative analysis, presented below.  \par
We estimate the size of the structures simply from the volume $V$ as:
\begin{equation}
R = V^{1/3}.  
\end{equation}
We define $N_{\rm tot}$ as the total number of morphologically classified structures, i.e. $N_{\rm tot}$ is
\begin{equation}\label{eq:ntot}
N_{\rm tot} = N_{\rm sheet} + N_{\rm sheet\_c} + N_{\rm filament} + N_{\rm spheroid},   
\end{equation}
with $N_x$ being the total number of structures (i.e. both parents and leaves) of morphological class $x$ (where $x\in$ [sheet, sheet\_c, filament, spheroid]). We express the number of structures of type $x$ at a given size $R$ by $N_x(R)$.

\par

\begin{figure*}
    \centering
    \includegraphics[width=0.49\textwidth]{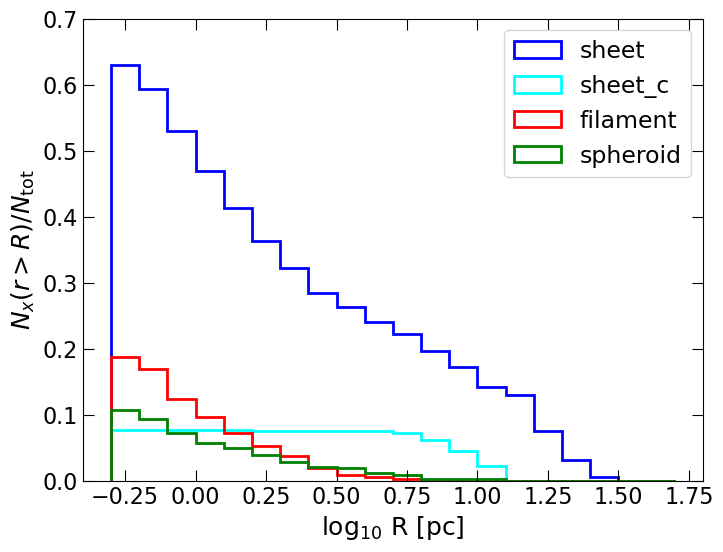}
    \includegraphics[width=0.49\textwidth]{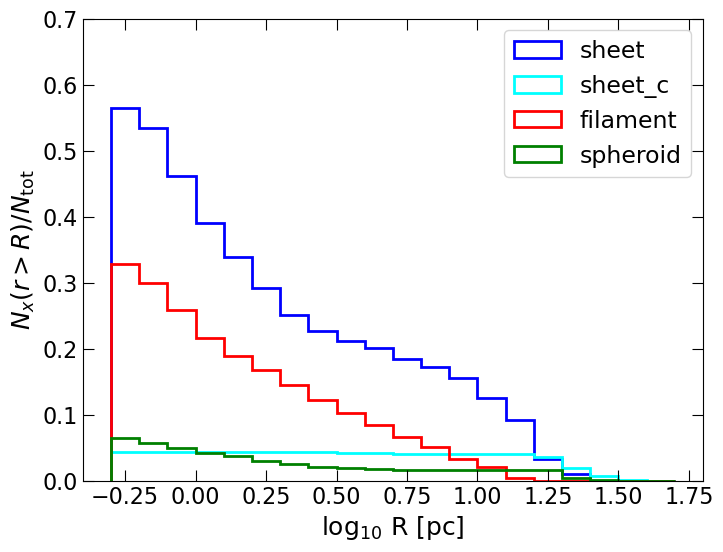}
    \caption{Cumulative histogram of different morphologies (sheets, curved sheets, filaments, or spheroids) for all HD (left) and MHD (right) clouds at $t_{\rm evol}=$~3.5~Myr. 6 out of the 7 analysed clouds are sheet-like on large scales, with filamentary networks embedded inside. Spheroidal structures are rarer in the presence of magnetic fields. Both HD and MHD clouds produce more sheets than filaments, but the MHD runs tend to have a relative increase in the fraction of filaments.}
    \label{fig:morphology}
\end{figure*}

In Fig.~\ref{fig:morphology} we plot the cumulative fraction (i.e. $N_x(R)/N_{\rm tot}$) of sheets, curved sheets, filaments, and spheroidal structures against $R$ for all structures (i.e. both parents and leaves) in the two hydrodynamic clouds (left panel) and the five MHD clouds (right panel) at $t_{\rm evol}=3.5$~Myr. The numerical values of the overall fractions across all scales, $\int N_x(R)/N_{\rm tot}\ \mathrm{d}R$, for both HD and MHD clouds at two different times can be found in Table~\ref{tab:morphology}. \par 

\begin{table}
    \centering
    \begin{tabular}{cccccc}
         \hline
         cloud, time&$\frac{N_{\rm sheet}}{N_{\rm tot}}$&$\frac{N_{\rm sheet\_c}}{N_{\rm tot}}$&$\frac{N_{\rm filament}}{N_{\rm tot}}$& $\frac{N_{\rm spheroid}}{N_{\rm tot}}$&$N_{\rm tot}$  \\
         \hline
         HD, 2 Myr&0.58&0.12&0.22&0.08&910\\
         HD, 3.5 Myr&0.63&0.07&0.19&0.11&1167\\
         \hline
         MHD, 2 Myr&0.57&0.03&0.31&0.09&487\\
         MHD, 3.5 Myr&0.56&0.04&0.33&0.07&2087\\
         \hline
    \end{tabular}
    \caption{Fraction of sheets, curved sheets, filaments, and spheroids among all morphologically classified structures, for both HD and MHD clouds at $t_{\rm evol}=2,\ 3.5$~Myr. While all clouds are dominated by sheet-like structures, the MHD clouds have a higher fraction of filaments compared to their hydrodynamic counterparts.}
    \label{tab:morphology}
\end{table}
We find that spheroidal structures, shown in green, are generally less numerous compared to sheet-like or filamentary structures ($\sim$10\% of $N_{\rm tot}$ are spheroidal, Table~\ref{tab:morphology}). Sheets (including curved sheets) appear to be the most abundant structures within all clouds (summing up to $\sim$70\% for the HD case and $\sim 60$\% for the MHD case). However, filaments are considerably more abundant in the MHD clouds compared to their HD counterparts ($> 30$\% for MHD as opposed to $\sim$20\% for HD clouds).  In terms of size, we find that at the largest $R$ values, indeed almost all clouds (six out of seven) are either sheets or curved sheets, confirming the visual trend we found in Fig.~\ref{fig:3d_projection}.
\par
\begin{figure*}
    \centering
    \includegraphics[width=0.49\textwidth]{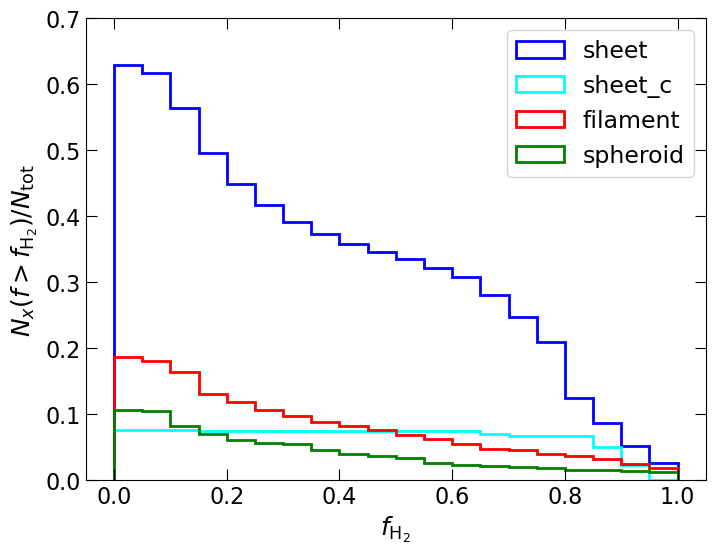}
    \includegraphics[width=0.49\textwidth]{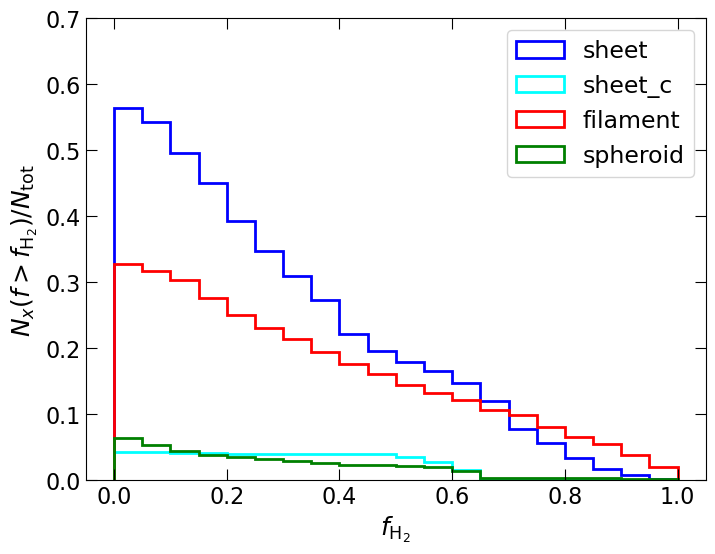}
    \caption{Cumulative histogram of different morphologies (sheets, curved sheets, filaments, or spheroids) against H$_2$ mass fraction for all HD (left) and MHD (right) clouds at $t_{\rm evol}=$~3.5~Myr. The most molecular structures are more filamentary in presence of magnetic fields.}
    \label{fig:morphology_mol_frac}
\end{figure*}

We highlight the morphological trends as a function of the molecular fraction in Fig.~\ref{fig:morphology_mol_frac}. Similar to Fig.~\ref{fig:morphology}, we plot here the cumulative fraction of (curved) sheets, filaments, and spheroids, but this time as a function of the molecular mass fraction $f_{\rm H_2}$, which is the H$_2$ mass in a structure divided by the total hydrogen mass in the structure. Note that structures with high $f_{\rm H_2}$ are usually small (located mostly at small $R$ in Fig.~\ref{fig:morphology}).
We see that around $f_{\rm H_2} > 0.7$, there are more filaments than sheet-like structures in the MHD case (right panel).
This trend is absent for the HD clouds (left panel). This implies that magnetic fields particularly enhance the formation of filaments on the small scales,  shaping the morphology of the denser, well-shielded, molecular gas. This is in line with the fact that magnetic fields can, in general, aid the formation of filamentary sub-structures \citep[][]{hacar_initial_2022, pineda_bubbles_2022}. \par
Gravitational collapse naturally proceeds anisotropically and tends to create elongated structures \citep[e.g.][]{burkert_collapse_2004}. However, we show in \citet{ganguly_silcc-zoom_2022} that most of our cloud structures are unbound or only marginally bound. This being the case, gravity cannot be the principal contributor to forming elongated structures, and we must therefore identify other possible sources of the lack of spheroidal structures. Shock compression and turbulence are two such methods for producing elongated structures (see, e.g., \citealt{inoue_formation_2016} for shock compression; \citealt{federrath_universality_2016} for turbulence; and \citealt{hacar_initial_2022} for a general overview).
Sheets and filaments are both elongated structures. However, it is interesting that for the hydrodynamic clouds, sheets are by far the most numerous, whereas for the MHD clouds filaments and sheets are more comparable in total number. This is consistent with the results of \citet{hennebelle_origin_2013}, who investigate setups of both decaying supersonic turbulence and colliding flows, and find that their simulations tend to produce more sheet-like structures for hydrodynamical simulations, and more filamentary structures for MHD simulations. \par
Overall, we see primarily sheet-like MCs with an abundance of elongated structures (filamentary or sheet-like), irrespective of whether the simulation contains magnetic fields or not. Sheets are generally more numerous, probably representing the fact that we trace a large number of structures belonging to the sheet-like atomic envelope of the MCs. This is supported by the fact that, in Fig.~\ref{fig:morphology_mol_frac}, both HD and MHD clouds show an abundance of sheet-like structures below $f_{\rm H_2} \approx 0.5$. The presence of magnetic fields, however, tends to somewhat increase the fraction of filamentary over sheet-like structures. \par
The sheet-like nature of our clouds is consistent with a number of recent observations. \citet{kalberla_cold_2016-1} have argued that the cold, neutral hydrogen in the ISM is organised in sheet-like structures. Investigating the L1495 region of the Taurus molecular cloud, \citet{arzoumanian_molecular_2018} report evidence of extended sheet-like structures too. Using the recent GAIA data, \citet{rezaei_kh_three-dimensional_2022} have concluded that the California molecular cloud is sheet-like in nature. \citet{tritsis_musca_2022} have reached a similar conclusion regarding the Musca molecular cloud using 3D dust extinction maps. Based on a Herschel study of the giant molecular filament G214.5, \citet{clarke_herschel_2023} have also posited that the filament is a result of the HI shell of an expanding superbubble interacting with the local medium.  Our findings here are thus perfectly in line with these observations.\par

The morphology of MCs at larger (tens of parsecs) scales is of paramount importance in relation to how the MCs themselves form. Our analysis shows that the clouds are preferentially sheet-like, with and without magnetic fields. The ISM in the SILCC simulations (and therefore also in the SILCC-Zoom simulations) has a multi-phase structure \citep[][]{walch_silcc_2015, girichidis_silcc_2016}. The MCs in these simulations form primarily at the shells or intersections of expanding supernova bubbles. The large-scale sheets we see, can therefore be interpreted as tracing these supernova-driven shells, with a complex network of different morphological sub-structure contained within. This picture is consistent with the bubble-driven structure formation scenario \citep[][]{koyama_molecular_2000,inoue_two-fluid_2009,inutsuka_formation_2015, pineda_bubbles_2022}. 

\section{Dynamics and Fragmentation}\label{sec:dynamics_and_fragmentation}\subsection{The magnetic field - density scaling}
The impact of magnetic fields on the MCs is naturally correlated to the field strength. The initial 3 $\mu$G seed field in the original simulations is expected to be enhanced when we look at denser structures inside the MCs. The scaling behaviour of the magnetic field $B$ with $\rho$ is integral to understanding the importance of magnetic fields at different scales. \par
If contraction of gas occurs exclusively along the magnetic field lines, this should lead to no dependence of the magnetic field strength on the density, i.e. $B\propto \rho^0$. If magnetic field lines do contract with the enhancement of gas density, then one expects a scaling similar to $B\propto \rho^{\kappa}$, with $\kappa=0.5,0.67$ for the strong and weak field limits, respectively \citep[see e.g. the review by][]{hennebelle_role_2019}. \par
In the ISM, indeed the $\kappa=0$ relation is observed up to number densities of $\sim$300 cm$^{-3}$ \citep[][]{troland_interstellar_1986, crutcher_magnetic_2010}. This corresponds to densities of roughly 1.1$\times10^{-21}$~g~cm$^{-3}$, using a mean molecular weight of 2.35. \citet{crutcher_magnetic_2010} find that above these densities, the data is consistent with $\kappa=2/3$, with considerable scatter. The transition in power law is usually associated with the magnetic fields becoming dynamically sub-dominant \citep[][]{seifried_parallel_2020, pattle_magnetic_2022} and roughly matches with our observation that below $\sim 100$~cm$^{-3}$ the mass in the MHD clouds is enhanced. \par  
We can attempt to capture whether this transition in the importance of the magnetic field is seen in the Alfv\'enic Mach number, $\mathcal{M}_{\rm A}$. For a given sub-structure, we can compute $\mathcal{M}_{\rm A}$ as 
\begin{equation}\label{eq:alfven_mach}
    \mathcal{M}_{\rm A}=\sigma_{\rm 1D}/v_{\rm A}.
\end{equation}
Here $\sigma_{\rm 1D}$ is the one-dimensional velocity dispersion and $v_{\rm A}$ is an estimate of the average Alfv\'en wave group velocity. For a structure of mass $M$, we compute $\sigma_{\rm 1D}$ from
\begin{equation}\label{eq:velocity_dispersion}
    \sigma_{\rm 1D}^2 = \frac{1}{3M}\int_V \rho (\mathbf{v} - \mathbf{v}_0)^2 \mathrm{d}^3r,
\end{equation}
with $\mathbf{v}_0$ being the centre of mass velocity computed as 
\begin{equation}\label{eq:bulk_velocity}
 \mathbf{v}_0 = \frac{1}{M}\int_V \rho \mathbf{v} \mathrm{d}^3r.  
\end{equation}
The integration is performed over the entire volume $V$ of the given structure. \par
The Alfv\'en velocity can be computed as 
\begin{equation}\label{eq:alfven}
    v_{\rm A} = \sqrt{\frac{\langle |\mathbf{B}|^2 \rangle}{4\pi\rho_{\rm avg}}}.
\end{equation}
The density $\rho_{\rm avg}$ here is the volume-averaged density, i.e. 
\begin{equation}\label{eq:density}
    \rho_{\rm avg}=M/V,
\end{equation} 
and $\langle |\mathbf{B}|^2 \rangle$ is the volume-averaged square of the magnetic field $\mathbf{B}$,
\begin{equation}
    \langle |\mathbf{B}|^2 \rangle = \frac{1}{V}\int_V |\mathbf{B}|^2 \mathrm{d}^3r.
\end{equation}\par

\begin{figure}
    \centering
    \includegraphics[width=0.49\textwidth]{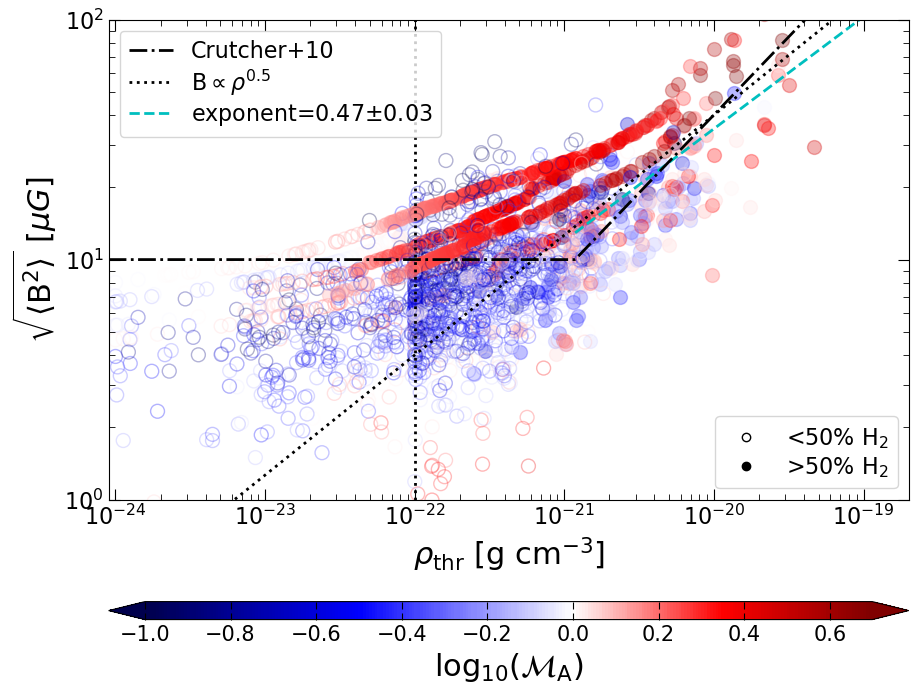}
    \caption{Relation between the root-mean-square magnetic field and $\rho_{\rm thr}$ for all MHD clouds at $t_{\rm evol}$=3.5~Myr. The colour bar shows the Alfv\'enic Mach number $\mathcal{M}_{\rm A}$. The dash-dotted line represents the B$-\rho$ relation from \citet{crutcher_magnetic_2010}, while the dotted line represents a $B\propto \rho^{0.5}$ power law. The cyan dashed line represents the best fit power law for all points with $\rho_{\rm thr}>1.1 \times 10^{-21}$~g~cm$^{-3}$. 
    }
    \label{fig:bn_relation}
\end{figure}

The behaviour of the magnetic field strength with density for the MHD clouds can be seen in Fig. \ref{fig:bn_relation}, where we plot the root-mean-square magnetic field strength against the threshold (minimum) density $\rho_{\rm thr}$ for all dendrogram structures at $t_{\rm evol}=3.5$~Myr. The different dendrogram structures are marked with filled/empty symbols depending on whether their H$_2$ mass fraction (with respect to their total hydrogen mass) is greater/less than 50\%. The colour bar shows $\mathcal{M}_{\rm A}$, as computed from Eq.~\ref{eq:alfven_mach}. The reddish points represent super-Alfv\'enic ($\mathcal{M}_{\rm A}>1$) structures, while the blueish points are sub-Alfv\'enic ($\mathcal{M}_{\rm A}<1$). In the sub-Alfv\'enic case, the fluid speed is smaller than the magnetic wave speed, meaning that the magnetic field is dynamically important and guides the flow.
The vertical dotted line at $10^{-22}$~g~cm$^{-3}$ represents the boundary between the points obtained from the \textit{low-den} (left half) and \textit{high-den} (right half) dendrograms, respectively. The dash-dotted black line and the dotted power-law represent the \citet{crutcher_magnetic_2010} relation discussed previously and $B\propto \rho^{0.5}$, respectively. \par

The cyan dashed line represents the linear least-squares best fit performed on the logarithm of the points for high densities ($\rho_{\rm thr}>1.1\times 10^{-21}$~g~cm$^{-3}$).
The best fit of $\kappa=0.47 \pm 0.03$ is consistent with the strong-field limit of $B\propto \rho^{0.5}$. 
We have already shown in the previous section (Section~\ref{sec:morphology}) that our structures are on average highly elongated, and magnetic fields clearly help to deform the shape of the forming structures. It is therefore not unexpected that we find a shallower scaling compared to the weak field limit ($\kappa=0.67$). \par
We see that, while there is no clear transition from the sub- to the super-Alfv\'enic regime, there is clearly a trend that higher Alfv\'enic Mach numbers are preferentially obtained at the higher density end. This is confirmed by a Kolmogorov-Smirnov (KS) two-sample test, which compares if two distributions belong to the same population. In this case, we compare the $\rho_{\rm thr}$-distributions of structures with $\mathcal{M}_A>1$ and $\mathcal{M}_A\leq1$. We find the $p$-values\footnote{If the $p$-value is larger than a certain value (typically 0.05), this means that we cannot reject the null hypothesis that the sub-Alfv\'enic and super-Alfv\'enic structures have the same underlying density distribution.} to be very low: $6\times 10^{-4}$ at 2 Myr and $5.2\times10^{-15}$ at 3.5 Myr (see Table~\ref{tab:ks_test}).\par  
\citet{crutcher_magnetic_2010} found that the observed magnetic field distribution is rather flat at low density, in agreement with the idea that denser clouds are swept up along the magnetic field lines on large scales, while at higher density there is a power-law increase of the magnetic field strength. If spherical clouds start to collapse and the magnetic field is not strong enough to stop the collapse, one expects a power-law slope of $\kappa=0.5 - 0.67$ (see above). \par 
In the case of our clouds, we find that the high-density end is well consistent with $\kappa=0.5$, and the lower-density end clearly shows a much shallower slope. Nonetheless, there does not seem to be a clear single density at which there is a sharp change in slope. Simulations by \citet{li_magnetized_2015}, \citet{mocz_moving-mesh_2017}, \citet{girichidis_magnetic_2018}, \citet{zhang_anchoring_2019} find similarly the lack of a sharp transition density. \citet{auddy_magnetic_2022} predict that the transition density depends on the fourth power of $\mathcal{M}_A$. While of potential interest, this is unfortunately not demonstrable from the present analysis. \par
\begin{table}
    \centering
    \begin{tabular}{c|c|c|c}
    \hline
         variable 1 & variable 2 &time [Myr] & p-value \\
         \hline
        $\rho_{\rm thr}(\mathcal{M}_A>1)$&$\rho_{\rm thr}(\mathcal{M}_A\leq1)$&2&$6\times 10^{-4}$\\
        &&3.5&$5.2\times10^{-15}$\\
        \hline
    \end{tabular}
    \caption{The $p$-values of the 2-sample KS test for the density distribution of sub-Alfvenic and super-Alfv\'enic structures. We can see that the $p$-value is low for both 2 and 3.5 Myr, suggesting that sub-Alfv\'enic and super-Alfv\'enic structures (corresponding to bluish and reddish points in Fig.~\ref{fig:bn_relation}, respectively) have statistically significant differences in their density distributions. }
    \label{tab:ks_test}
\end{table}

\subsection{Impact of magnetic fields on the energetics of sub-structures}\label{sec:dynamics}
We are also interested in assessing the energetic relevance of magnetic fields over different length scales in the MCs, especially with respect to potentially star-forming structures. For this purpose, we compute the volume term of the magnetic energy and compare it with the kinetic and potential energies. Similar work for the same simulations has been performed by \citet{ganguly_silcc-zoom_2022}, who assess the virial balance of the cloud sub-structures. 
Here, we extend the range of our analysis to include the dynamics of lower-density gas (between 10$^{-24}$ and 10$^{-22}$ g~cm$^{-3}$; \textit{low-den} dendrogram analysis, see Table~\ref{tab:dendrogram_type}). \par 
The magnetic energy of a given structure is computed as
\begin{equation}
    E_{\rm B} = \int_V \frac{1}{8 \pi} |\mathbf{B}|^2 \mathrm{d}^3r,
\end{equation}
where the integration is computed over the entire volume $V$ of the structure. The kinetic energy is computed using the following relation:
\begin{equation}
    E_{\mathrm{KE}} = \frac{1}{2}\int_V\rho(\mathbf{v}-\mathbf{v}_0)^2\mathrm{d}^3r.
\end{equation}\label{eq:ke}
\par 
Here, $\mathbf{v}_0$ is the centre of mass velocity computed from Eq.~\ref{eq:bulk_velocity}. The self-gravitating potential energy of a given structure is obtained using the following relation:
\begin{equation}
    E_{\rm PE} = -\frac{1}{2}G\int_V\int_V \frac{\rho(\mathbf{r}) \rho(\mathbf{r'})}{|\mathbf{r}-\mathbf{r'}|} \mathrm{d}^3r \mathrm{d}^3r',
\end{equation}
where $G$ is the gravitational constant. We compute the self-gravity of each dendrogram structure using a KD-tree algorithm \citep[][]{bentley_multidimensional_1975} instead of an $\mathcal{O}(N^2)$ direct computation.\par
\begin{figure*}
    \centering
    \includegraphics[width=0.49\textwidth]{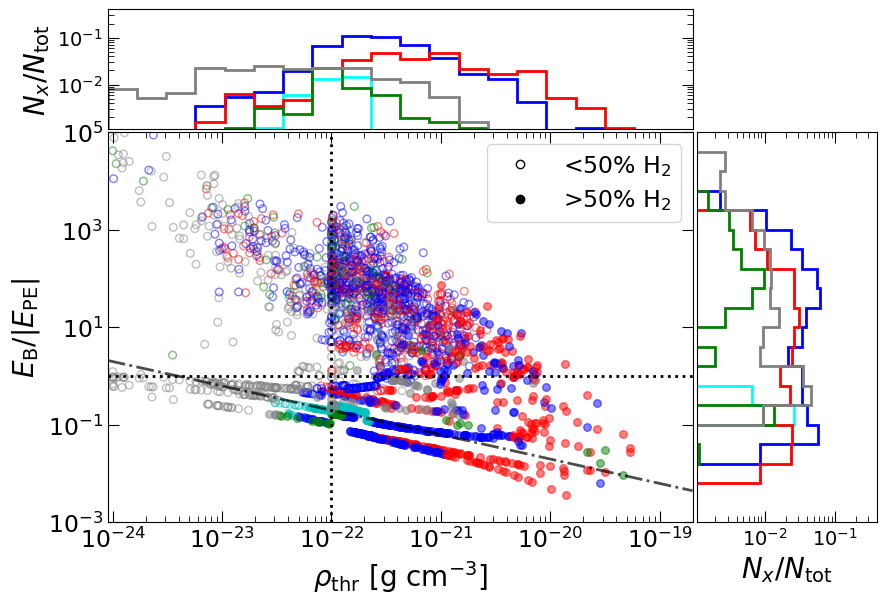}
    \includegraphics[width=0.49\textwidth]{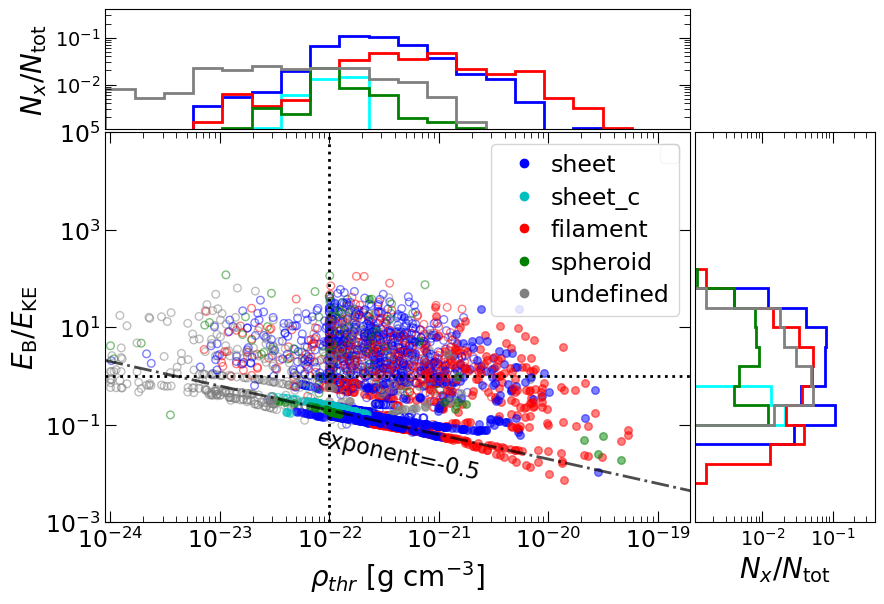}

    \caption{Ratio of magnetic energy to self-gravitating potential energy (left) and to kinetic energy (right), respectively, plotted against the density threshold for all dendrogram structures of all MHD clouds at time $t_{\rm evol} = 3.5$~Myr. The colours represent different morphologies. The dash-dotted lines indicate a $\rho^{-1/2}$ relation. The top and the right panels show the marginalised distributions (separated by morphology) over the density and the corresponding energy ratio.}
    \label{fig:energetics}
\end{figure*}

We show the relative importance of magnetic fields with respect to potential and kinetic energy in the left and right panel of Fig. \ref{fig:energetics}, respectively, for all MHD cloud structures at $t_{\rm evol}=3.5$~Myr. For both plots, the $x$-axis represents the density threshold $\rho_{\rm thr}$, and the $y$-axis represents $E_{\rm B}/|E_{\rm PE}|$ (left) and $E_{\rm B}/|E_{\rm KE}|$ (right), respectively. The colours of the points represent their morphologies. Here, for the purpose of understanding the dynamics of low-density gas, we also include the "unclassified" structures (i.e. structures with >5\% of their surface cells touching the edge of the analysis box, see Section~\ref{sec:classification}). 
The side panels to the right and top of each plot show the marginal distributions of $N_x/N_{\rm tot}$ for each morphology. Note that, since the definition of $N_{\rm tot}$ (Eq.~\ref{eq:ntot}) does not contain unclassified structures, the fractions in the two side panels add up to greater than unity. The filled symbols are molecular structures, while the open symbols are atomic.
Typically, for low-density structures, which mostly consist of atomic gas, the magnetic energy is either comparable to or much larger than the potential energy (left panel of Fig.~\ref{fig:energetics}). The magnetic energy is also comparable to or larger than the kinetic energy (right panel), but the spread in this energy ratio is much smaller compared to the $E_{\rm B}/E_{\rm PE}$ ratio.
For some branches (a dendrogram branch is defined as a given structure and all its parent structures, see Section~\ref{sec:methods_structure_identification}), the energy ratio seems to roughly follow a $\rho^{-1/2}$ power law. These branches represent the evolution from diffuse, large-scale structures to denser, embedded structures. \citet{camacho_kinetic_2022} also find a tight power-law scaling between the potential and magnetic energies. While not exactly the same, both scaling behaviours seem to imply that magnetic fields become less important as we go deeper into the MCs themselves. This is also in accordance with the findings of \citet{seifried_parallel_2020}, \citet{ibanez-mejia_gravity_2022}, as well as \citet{ganguly_silcc-zoom_2022}, as discussed previously. \par
From the marginal distributions, we find a weak trend that the high-density end is dominated by filaments. Curved sheets and unclassified structures only appear at lower densities because they are usually larger-scale structures. There is no obvious correlation between the morphology of the structures and the energy ratios. This suggests that the different morphological configurations are created by the same formation mechanism, most likely turbulent compression. \par

\begin{figure*}
    \centering
    \includegraphics[width=0.49\textwidth]{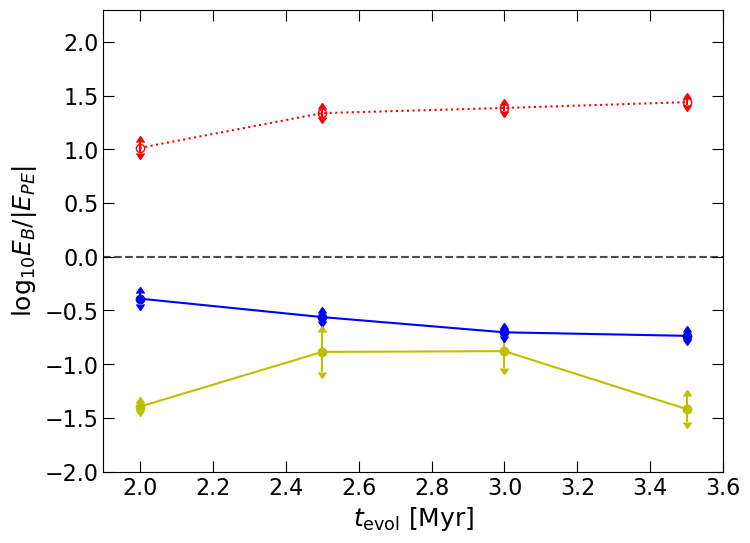}
    \includegraphics[width=0.49\textwidth]{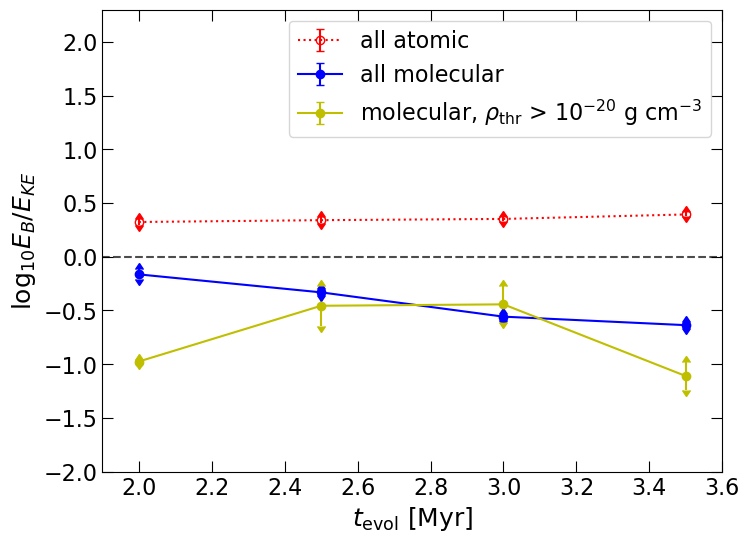}
    \caption{Time evolution of the average ratio of magnetic to potential energy (left) and kinetic energy (right). The different colours represent atomic, molecular, and dense molecular (molecular and $\rho_{\rm thr}>10^{-20}$~g~cm$^{-3}$) structures in red, blue, and yellow, respectively. The errors bars are the standard errors on the mean. For denser and molecular structures, magnetic energy is less important compared to potential or kinetic energies. The atomic structures, representing more the envelope of the molecular gas, have high magnetic energies, especially compared to self-gravity.}
    \label{fig:avg_evolution}
\end{figure*}

There also seems to be a difference in the energy ratios between atomic and molecular structures. This can be clearly seen in the average behaviour of these ratios over time. Fig. \ref{fig:avg_evolution} plots the time evolution of the average value of $E_{\rm B}/|E_{\rm PE}|$ (left) and $E_{\rm B}/E_{\rm KE}$ (right) for all atomic (red), molecular (blue), and dense molecular (yellow) structures from the MHD clouds, where we define dense molecular structures to be structures that are both molecular and have $\rho_{\rm thr}>10^{-20}$~g~cm$^{-3}$. The error bars here represent the standard error on the mean. From Fig.~\ref{fig:avg_evolution}, 
we see that magnetic energy dominates over potential and kinetic energies for atomic structures, while it plays a subordinate role in molecular structures. There is no clear trend indicating that this behaviour changes as a function of time.\par

The subservient role of magnetic energy for dense structures compared to potential or kinetic energy suggests that while magnetic fields help to shape the cloud structures across different scales, the dynamics of the denser, and potentially star-forming structures, is determined by the interaction between gravity and turbulence \footnote{We explore the interplay between turbulence and gravity in much greater detail in our companion paper by means of a virial analysis \citep{ganguly_silcc-zoom_2022}}. This explains why there is no discernible difference in the power-law tail of the density PDFs between hydrodynamic and MHD clouds (Fig.~\ref{fig:density_pdf}), confirming that the star-forming gas \citep[see e.g.][]{klessen_formation_2001, girichidis_evolution_2014, schneider_understanding_2015} is virtually unaffected by the presence of magnetic fields. However, magnetic fields change the gas properties of the environment from which denser structures form, accrete, and sit in (i.e. by making the surrounding envelope "fluffier"), thereby also influencing the shape of these structures.
 \par 

\subsection{Magnetic surface energy}\label{sec:bsurface}
In the previous section, we have discussed the magnetic pressure term in comparison to self-gravity and kinetic energy. The magnetic pressure relates to the stretching and compression of magnetic field lines, and does not take into account the effect of curvature in the field. \par
The magnetic surface term can be computed as an integral over the surface of a given structure, $S$, as follows:
\begin{equation}\label{eq:bsurface}
    E_{\rm B}^{\rm surface}= \oint_S (\mathbf{r}-\mathbf{r}_0) \mathbf{T} \hat{\mathbf{n}}\ \mathrm{d}S.
\end{equation}
Here $\mathbf{r}_0$ is the centre of mass, $\hat{\mathbf{n}}$ is the surface normal vector that points outwards, and $\mathbf{T}$ is the Maxwell stress tensor, which can be written as follows for ideal MHD:
\begin{equation}
    \mathbf{T} = \frac{1}{4\pi}\left(\mathbf{B}\otimes\mathbf{B}-\frac{1}{2}|\mathbf{B}|^2\hat{\mathbf{I}}\right).
\end{equation}
$\hat{\mathbf{I}}$ is here an identity matrix of rank two. \par 
We evaluate Eq.~\ref{eq:bsurface} as a volume integral using the Gauss' divergence theorem for convenience. This gives us the following relation:
\begin{equation}\label{eq:bsurface_vol}
    E_{\rm B}^{\rm surface}=- E_{\rm B}+\int_V (\mathbf{r}-\mathbf{r}_0)\cdot \nabla \mathbf{T}\ \mathrm{d}V
\end{equation}
From Eq. \ref{eq:bsurface_vol}, we can see that $E_{\rm B}^{\rm surface}$ can be both positive or negative. When it is is positive, it adds to the magnetic pressure term and acts as a dispersive term. In contrast, when $E_{\rm B}^{\rm surface}<0$, it acts as a confining term. \par
\begin{figure*}
    \centering
    \includegraphics[width=0.49\textwidth]{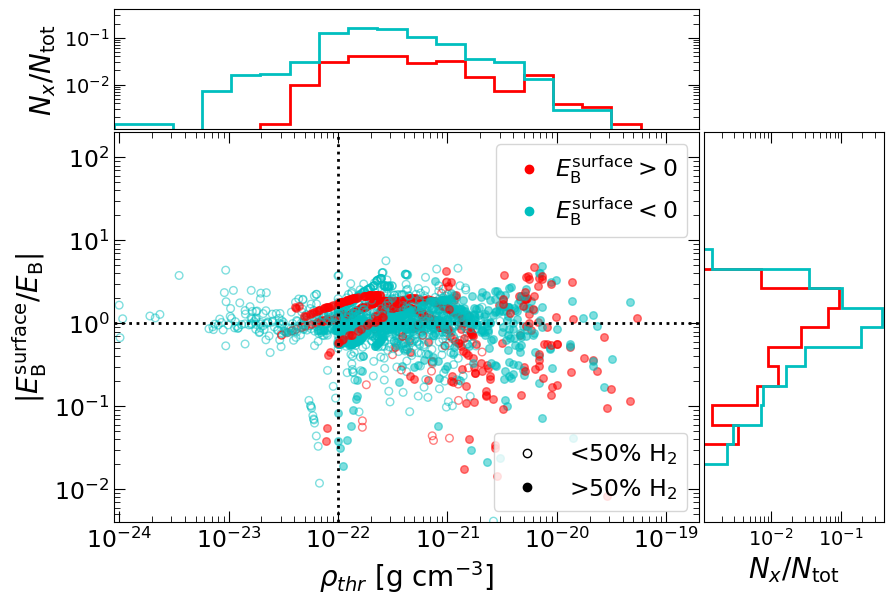}
    \includegraphics[width=0.49\textwidth]{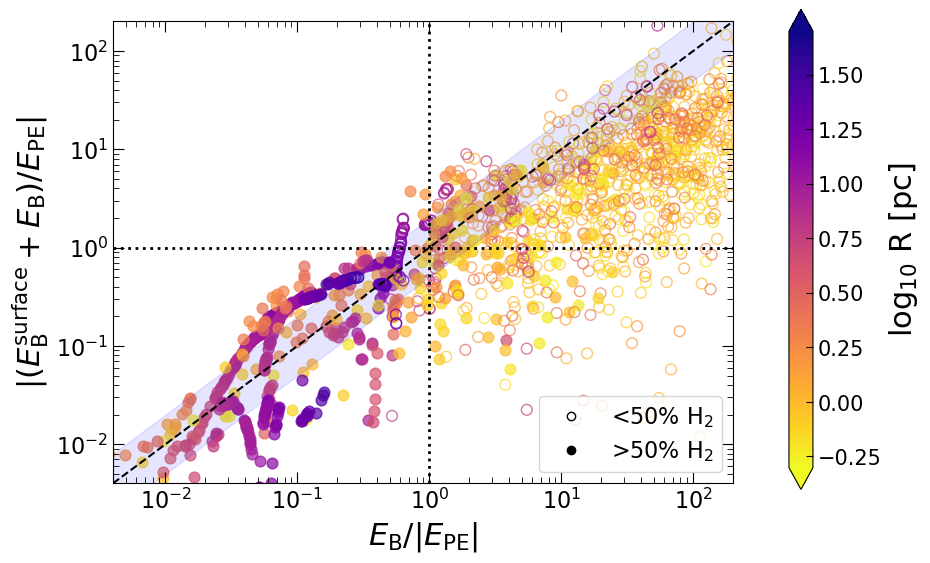}
    \caption{Left: Ratio of the absolute value of the magnetic surface to volume energy, plotted against the density threshold. The different colours represent whether the magnetic surface term is positive and resists collapse or negative and promotes collapse. The magnetic surface energy seems to be as relevant as the volume energy, and for more than half of the structures acts as a confining term. Right: The ratio of the total magnetic energy (surface plus volume) to the self-gravitating potential energy, plotted against the magnetic volume energy over the self-gravitating potential energy. The dashed line represents a 1:1 ratio, and the shaded region represents a factor of 2. For many small-scale atomic structures, the magnetic surface term seems to be important as a confining force.}
    \label{fig:magnetic_surface}
\end{figure*}

The importance of $E_{\rm B}^{\rm surface}$ with respect to the volume term, $E_{\rm B}$, can be seen in Fig. \ref{fig:magnetic_surface}, left panel, which plots the magnitude of the ratio of $E_{\rm B}^{\rm surface}/E_{\rm B}$ to the density threshold of the cloud sub-structures for all MHD clouds at $t_{\rm evol}=3.5$~Myr. Structures where $E_{\rm B}^{\rm surface}$ helps to disperse them ($E_{\rm B}^{\rm surface}>0$) are marked in red, while structures where $E_{\rm B}^{\rm surface}$ acts as a confining term ($E_{\rm B}^{\rm surface}<0$) are marked in cyan. The vertical dotted line marks the difference between the results of the \textit{low-den} and \textit{high-den} dendrogram runs at $\rho=10^{-22}$~g~cm$^{-3}$, as in the previous plots. The horizontal dotted line represents a value of one, where the volume and surface terms are equally important magnitude-wise. The top and side panels show the marginal distributions.
From the marginal distributions, we see that $E_{\rm B}^{\rm surface}$ acts as a confining term for somewhat more number of structures compared to where the surface term is dispersive. The magnetic surface term seems to be comparable to and in some cases, even exceeding the volume term $E_{\rm B}$. This implies that for diffuse and mostly atomic structures, where magnetic energy is comparable or dominant, the surface term is important. This is especially relevant when $E_{\rm B}^{\rm surface}$ acts as a confining term. However, for dense structures, where $E_{\rm B}$ is one to two orders of magnitude smaller than the potential and kinetic energies, the surface term is unlikely to significantly affect the dynamics. \par
In the right panel of Fig.~\ref{fig:magnetic_surface} we plot the magnitude of $(E_{\rm B}^{\rm surface}+E_{\rm B})/E_{\rm PE}$ against $E_{\rm B}/E_{\rm PE}$ for all MHD cloud sub-structures at 3.5~Myr. The colour bar here represents the size of the structures. The horizontal and vertical dotted lines both represent a value of unity along the $y-$ and $x-$ axes, respectively. The dashed line represents a 1:1 line, and the shaded region around it represents a factor of 2 in each direction. The magnetic surface energy is not significant compared to the volume energy for structures on or close to the 1:1 line. Structures with strong dispersive $E_{\rm B}^{\rm surface}$ terms lie above the 1:1 line, while points that lie below the 1:1 line represent structures where $E_{\rm B}^{\rm surface}$ is confining in nature. Most interesting here are the points that lie in the bottom right quadrant of the plot. They represent structures where the magnetic pressure $E_{\rm B}$ is higher compared to the self-gravity, and would be completely unbound in a traditional virial analysis. However, the confining $E_{\rm B}^{\rm surface}$ term is strong enough that the overall magnetic contribution becomes far less, thus allowing for a sort of "magnetic confinement". These structures are mostly atomic and typically seem to be $\lessapprox 1$~pc. \par
\begin{figure*}
    \centering
    \includegraphics[width=0.49\textwidth]{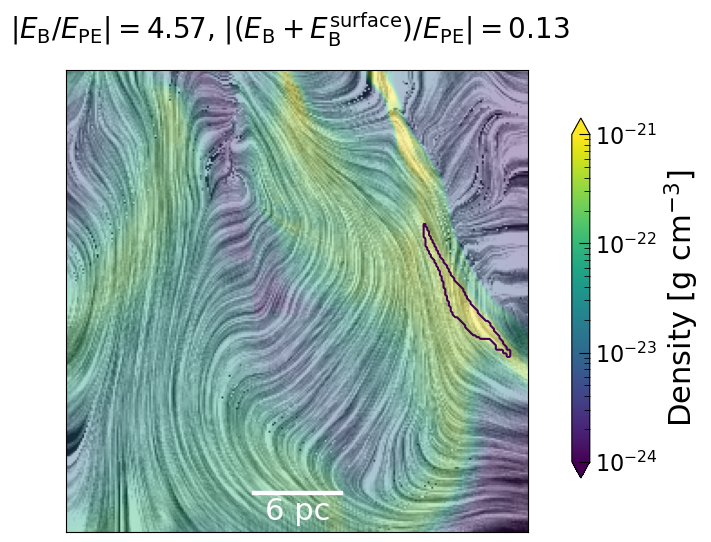}
    \includegraphics[width=0.49\textwidth]{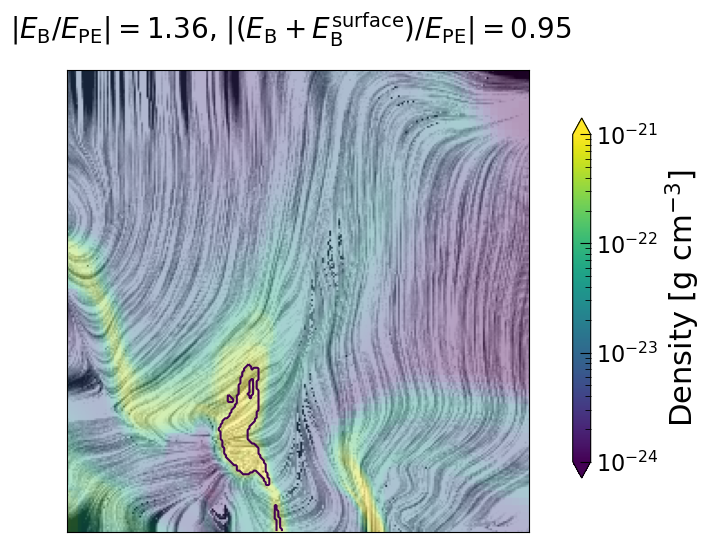}
    \caption{Two examples of structures confined by $E_{\rm B}^{\rm surface}$ from MC2-MHD, plotted as black contours over density slices in the $y-z$ plane, at $t_{\rm evol}=2$~Myr. The colour map is the logarithmic density, and the direction of the planar magnetic field is plotted as line integral convolution. The relevant energy ratios of the indicated structures are denoted in the title. Structures for which the magnetic surface energy is important and of confining nature, are usually located at the "kinks" of magnetic field lines.}
    \label{fig:magnetic_confnement}
\end{figure*}

 
Two examples of structures belonging to MC2-MHD that exhibit such magnetic confinement are plotted in Fig. \ref{fig:magnetic_confnement} as black contour lines over a density slice in the $y-z$ plane. The background colour here represents the density, while the planar magnetic field is shown using the line integral convolution (LIC) technique\footnote{The package used can be found in \url{https://github.com/alexus37/licplot}.}. For both structures, we mention the magnitude of the $E_{\rm B}/E_{\rm PE}$ and $(E_{\rm B}^{\rm surface}+E_{\rm B})/E_{\rm PE}$ ratios in the figure title. As can be clearly seen, the magnetic surface term reduces the $|(E_{\rm B}^{\rm surface}+E_{\rm B})/E_{\rm PE}|$ ratio to less than one. However, this naturally does not take into account other energy terms, i.e. kinetic and thermal energy, and hence it is not fully clear whether these structures are overall confined. Interestingly, the structures for which the magnetic surface energy is important and of confining nature (see right panel of Fig.~\ref{fig:magnetic_surface}) are usually located at the "kinks" of magnetic field lines. \par 

\subsection{Fragmentation}\label{sec:fragmentation}

In this section, we attempt to quantify to what extent magnetic fields affect the fragmentation properties of molecular clouds. For this purpose, we study the numbers and masses of different fragments, represented by leaf structures (i.e. structures containing no further sub-structures) found in our dendrogram analysis, and in addition perform a magnetic Jeans analysis on these fragments. \par
\subsubsection{Number and mass distribution of fragments}
Representing fragments by the leaves in the dendrogram analysis suffers from the caveat of depending on the dendrogram parameters. Increasing the minimum number of cells required in a dendrogram structure, for example, would naturally reduce the number of fragments and increase their masses. The absolute values of the masses and numbers we find, therefore, are sensitive to the parameter values we have used. However, since we used the exact same parameters for each HD and MHD run, and because all molecular clouds have similar masses and identical environmental parameters (solar neighbourhood parameters), the relative difference between the average behaviour of the HD and MHD clouds is meaningful. With this caveat in mind, let us look at the fragmentation properties of our dendrogram structures. \par 
\begin{figure*}
    \centering
    \includegraphics[width=0.33\textwidth]{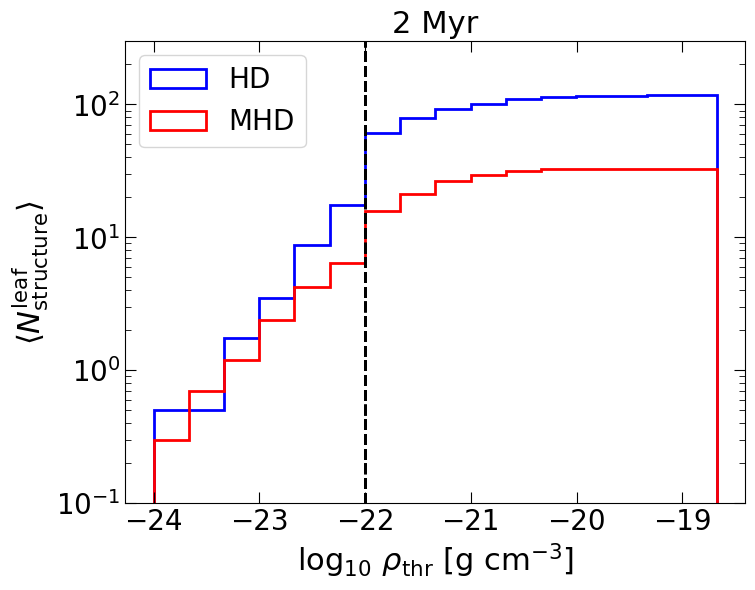}
    \includegraphics[width=0.33\textwidth]{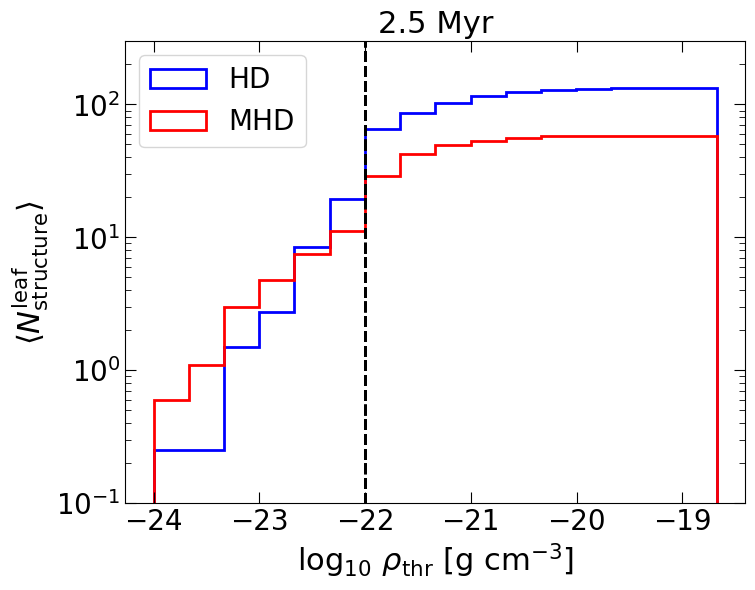}
    \includegraphics[width=0.33\textwidth]{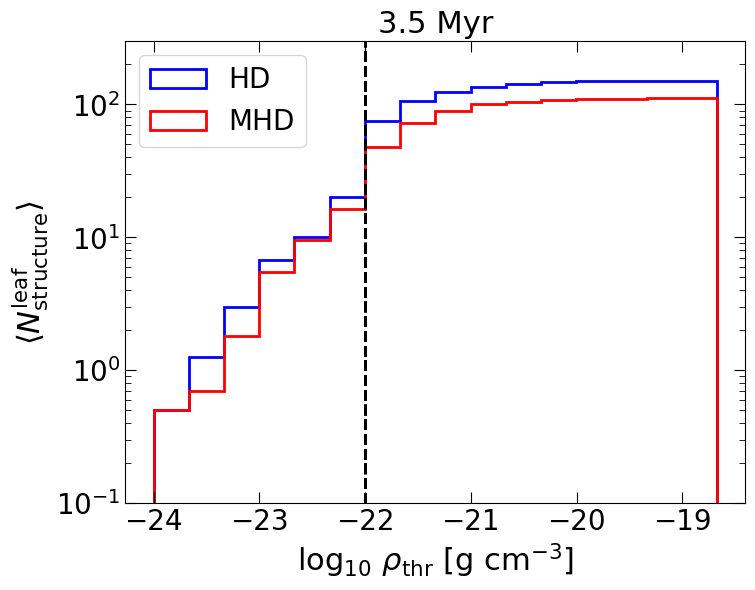}
    \includegraphics[width=0.33\textwidth]{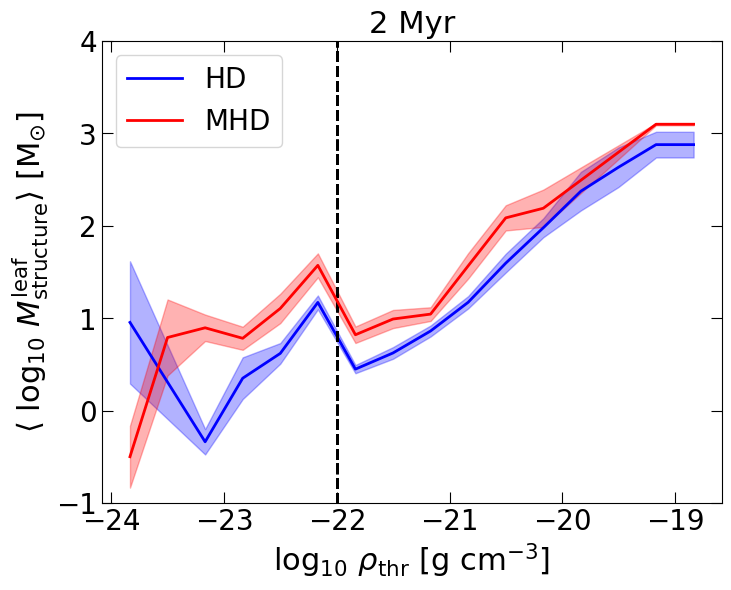}
    \includegraphics[width=0.33\textwidth]{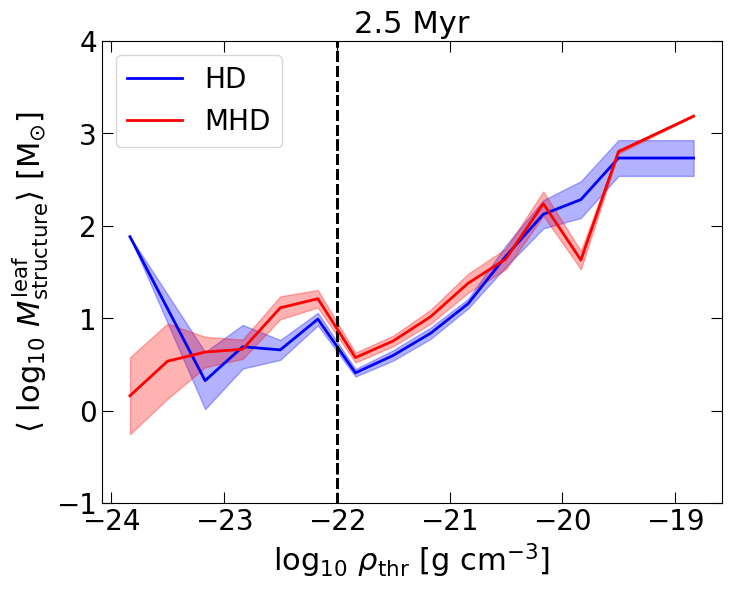}
    \includegraphics[width=0.33\textwidth]{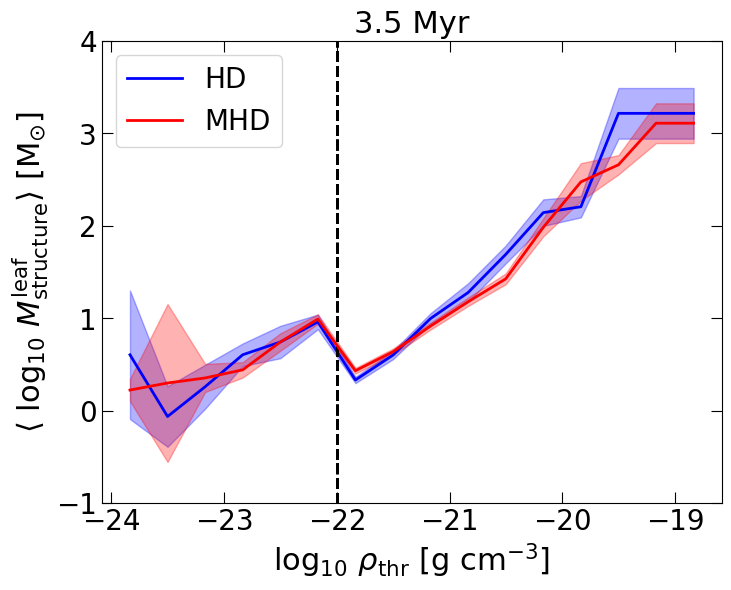}
    \caption{Top row: Cumulative distribution of the average number of leaf structures against $\rho_{\rm thr}$ for HD and MHD clouds at $t_{\rm evol}=$~2,~2.5,~3.5 Myr, respectively (from left to right). The hydrodynamic clouds have on average more new structures forming at earlier times, but this distinction slowly disappears later on. Bottom row: Distribution of average mass of leaf structures for both HD and MHD clouds at $t_{\rm evol}=$~2,~2.5,~3.5 Myr, respectively (from left to right). The leaf structures, representing fragments, are more massive for MHD clouds at earlier times, while this distinction mostly disappears later on as gravity takes over.}
    \label{fig:fragmentation}
\end{figure*}
We study the numbers and masses of leaf fragments in Fig. \ref{fig:fragmentation}. The top row plots the cumulative distribution of the average number of leaf structures, $\langle N_{\rm structure}^{\rm leaf}\rangle$, as a function of $\rho_{\rm thr}$ for both HD (blue) and MHD (red) clouds. The average here simply means that we divide the total number of obtained structures by the number of clouds, i.e. 5 for MHD and 2 for HD. The three panels (left to right) show three different times, $t_{\rm evol}=2,\ 2.5$ and 3.5~Myr, respectively. 
The vertical line at $10^{-22}$~g~cm$^{-3}$ marks the difference between the \textit{low-den} and the \textit{high-den} dendrogram analysis. 
We see that at $t_{\rm evol}=2$ and 2.5~Myr, up to densities between $10^{-23}-10^{-22}$~g~cm$^{-3}$, the HD and MHD clouds form roughly similar numbers of leaf fragments. However, at higher densities, $\langle N_{\rm structure}^{\rm leaf}\rangle$ is much higher for the HD clouds. This difference largely disappears at 3.5 Myr.
This suggests that the formation of structures is somewhat slowed down in the presence of magnetic fields in the beginning, but at later stages, as gravity becomes dynamically more and more important, this difference diminishes. \par

In the bottom row of Fig.~\ref{fig:fragmentation} we plot the average mass of the leaf structures, $\langle M_{\rm structure}^{\rm leaf} \rangle$, as a function $\rho_{\rm thr}$ for HD and MHD structures for the three different times. The shaded regions represent the standard deviation of the average mass at a given $\rho_{\rm thr}$.
We see that at $t_{\rm evol}=2$~Myr, the MHD fragments are slightly more massive compared to their hydrodynamic counterparts, in particular for $\rho_{\rm thr}\lesssim 10^{-21}$~g~cm$^{-3}$. This difference disappears later. For the densest structures, we do not seem to see a systematic difference in $\langle M_{\rm structure}^{\rm leaf} \rangle$. This is in line with Fig.~\ref{fig:density_pdf}, which shows that the difference in the density PDFs between the HD and MHD clouds in the density range that corresponds primarily to the cloud envelope (i.e. $\lessapprox 10^{-21}$~g~cm$^{-3}$) is most striking at $t_{\rm evol}=2$~Myr, and less so later on. \par
Overall, the results shown in Fig.~\ref{fig:fragmentation} indicate that the MHD clouds fragment more slowly than the HD clouds but therefore have slightly more massive fragments at early times. 
This is consistent with the result that magnetic fields affect the dynamics of lower density gas more \citep[][]{molina_density_2012, seifried_silcc-zoom_2020, seifried_parallel_2020, ibanez-mejia_gravity_2022}. We also see that the number and mass of the leaf structures are comparable at later times. This suggests that the magnetic fields "slow down" the evolution of the cloud but are less relevant once the cloud is more evolved, and gravity becomes energetically more and more important, as shown in the previous energetic analysis. This effect could be related to the overall strength of the magnetic field. We investigate this further in the next Section.

\subsubsection{Magnetic Jeans analysis}
The classic thermal Jeans analysis \citep[][]{jeans_stability_1902} is a useful tool to investigate the stability of MCs and their substructures (clumps and cores) under thermal perturbations. Here, we perform its magnetic equivalent. The thermal Jeans length, $\lambda_{\rm T}$, defines the largest length-scale stable to thermal perturbations. For a given structure, this is defined as
\begin{equation}
    \lambda_{\rm T} = c_s \sqrt{\frac{\pi}{G\rho_{\rm avg}}},
\end{equation}
where $c_s$ is the average sound speed given by
\begin{gather}
    c_s = \frac{1}{V}\int_V \sqrt{\frac{P}{\rho}} \mathrm{d}^3r.\label{eq:jeans_length_thermal}
\end{gather}
Here, $P$ is the thermal pressure and the sound speed is calculated assuming an isothermal equation of state due to the densities under consideration. We remind the reader that $\rho_{\rm avg}$ is the volume-averaged density computed in Eq.~\ref{eq:density}. From the Jeans length, a maximum mass stable under thermal perturbations can be calculated. This mass is referred to as the thermal Jeans mass, $M_{\rm T}$, and is given by
\begin{equation}
    M_{\rm T} = \frac{4}{3}\pi \rho_{\rm avg} \left(\frac{\lambda_{\rm T}}{2}\right)^3.
\end{equation}
\par
Similar to the thermal analysis, we can perform a magnetic Jeans analysis and a Jeans analysis combining both magnetic and thermal support. For the magnetic Jeans analysis, the relevant length ($\lambda_{\rm B}$) and mass ($M_{\rm B}$) scales are given by,
\begin{gather}
   \lambda_{\rm B} = c_{\rm B} \sqrt{\frac{\pi}{G\rho_{\rm avg}}},\\
    M_{\rm B} = \frac{4}{3}\pi \rho_{\rm avg} \left(\frac{\lambda_{\rm B}}{2}\right)^3.
\end{gather}
For a combination of thermal and magnetic effects, the relevant magneto-thermal Jeans length ($\lambda_{\rm B,T}$) and Jeans mass ($M_{\rm B,T}$) are
\begin{gather}
   \lambda_{\rm B,T} = c_{\rm B,T} \sqrt{\frac{\pi}{G\rho_{\rm avg}}},\\
    M_{\rm B,T} = \frac{4}{3}\pi \rho_{\rm avg} \left(\frac{\lambda_{\rm B,T}}{2}\right)^3.\label{eq:magnetic_thermal_jeans_mass}
\end{gather}
The characteristic speeds are given by,
\begin{gather}
    c_{\rm B} = v_{\rm A},\\
    c_{\rm B,T} = \sqrt{c_s^2+ v_{\rm A}^2},
\end{gather}
where $v_{\rm A}$ is the Alfv\'en speed (Eq.~\ref{eq:alfven}). \par

\begin{figure}
    \includegraphics[width=\columnwidth]{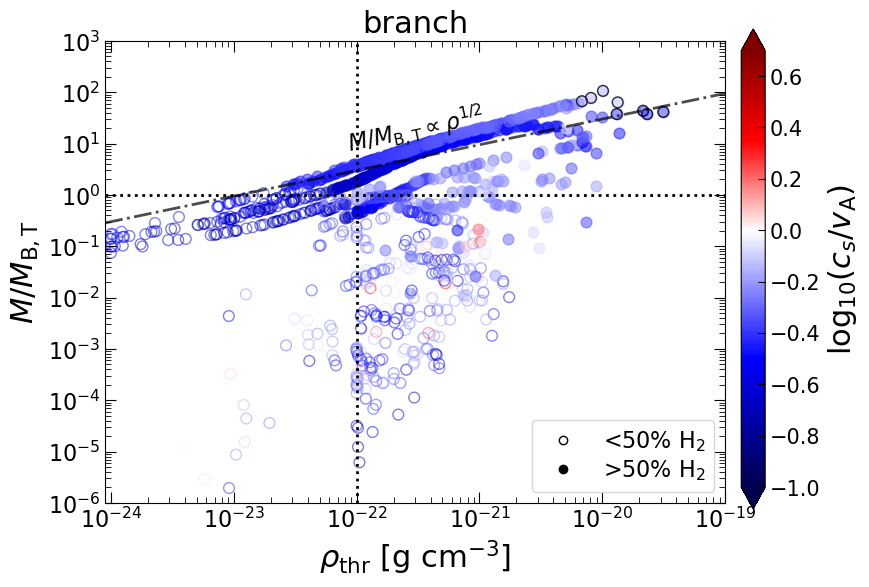}
    \includegraphics[width=\columnwidth]{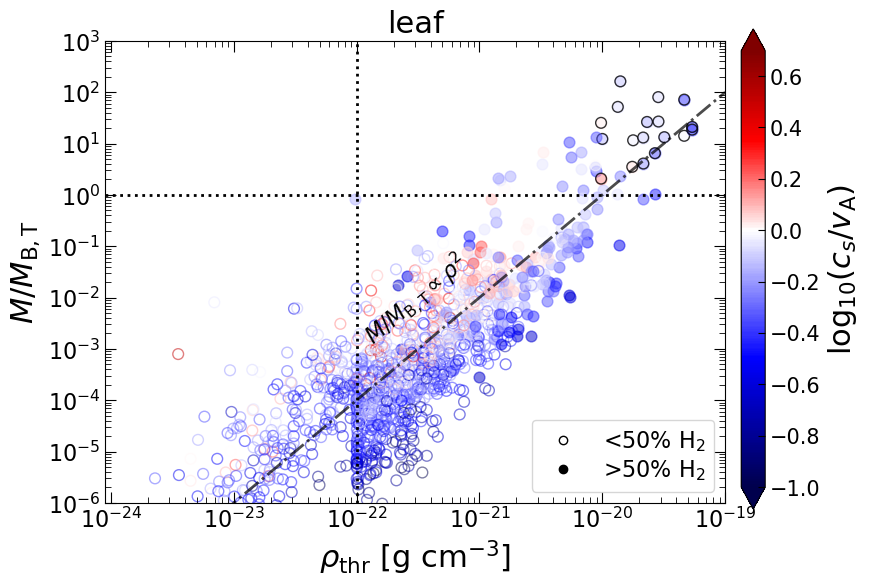}
    \caption{The ratio of the mass of a given structure to its magneto-thermal Jeans mass ($M_{\rm B,T}$, Eq.~\ref{eq:magnetic_thermal_jeans_mass}) as a function of $\rho_{\rm thr}$ for all MHD branch (top) and leaf (bottom) sub-structures at $t_{\rm evol}=3.5$~Myr. A branch sub-structure has further sub-structures, while a leaf does not. The horizontal dotted line represents a ratio of unity. The vertical line separates the points obtained from the \textit{high-den} and the \textit{low-den} dendrogram analysis. The colour-bar shows the ratio of the sound speed to the Alfv\'en wave speed. For blue points, $v_{\rm A}> c_s$. A power-law is plotted in each panel for rough guidance. Structures whose fragmentation is not well-resolved (see Fig.~\ref{fig:magnetic_thermal_jeans_length}) are marked with an additional black outline and mostly populate the right-hand top corner of the plot. The magneto-thermal forces seem unable to keep all the structures Jeans stable beyond $\sim 10^{-22}$~g~cm$^{-3}$, suggesting the growing importance of gravity.}
    \label{fig:magnetic_thermal_jeans}
\end{figure}

In Fig.~\ref{fig:magnetic_thermal_jeans}, we show the ratio of a structure's mass to its magneto-thermal Jeans mass, $M/M_{\rm B,T}$, as a function of $\rho_{\rm thr}$ for all MHD cloud branch structures (top) and leaves (bottom) at $t_{\rm evol}=3.5$~Myr. We remind the reader that branch structures contain sub-structures and leaves do not. The Jeans mass can only be properly used when the corresponding length is resolved. This is shown in Appendix~\ref{sec:app_magnetic_jeans}, Fig.~\ref{fig:magnetic_thermal_jeans_length}, which depicts that some structures with $\rho_{\rm thr} \gtrapprox 10^{-20}$~g~cm$^{-3}$ seem to be not properly Jeans resolved. These are marked with black outlines in Fig.~\ref{fig:magnetic_thermal_jeans}. Note that the structures are \textit{(un-)resolved} in the context of our dendrogram analysis, which requires a minimum number of 100 cells per structure, and therefore at least 200 cells to resolve fragmentation (as in the case of fragmentation, each fragment would need to contain at least 100 cells). 
The colour-bar denotes the ratio of $c_s$ to $v_{\rm A}$. Most of 
the structures have $v_{\rm A}> c_s$ (blue points), suggesting support by magnetic fields rather than by thermal pressure.
This is confirmed by our purely magnetic Jeans analysis (Fig.~\ref{fig:magnetic_jeans}), which shows an almost identical distribution to the magneto-thermal Jeans analysis of Fig.~\ref{fig:magnetic_thermal_jeans} (as well as Fig.~\ref{fig:magnetic_thermal_jeans_length}).
From Fig.~\ref{fig:magnetic_thermal_jeans}, top panel, we find that roughly below $10^{-22}$~g~cm$^{-3}$, all structures are Jeans stable ($M/M_{\rm B,T}<1$). At higher densities, we have both, Jeans stable and unstable structures. Some prominent branches clearly have $M/M_{\rm B,T}>1$ above $10^{-22}$~g~cm$^{-3}$, indicating the growing importance of gravity for fragmentation at higher densities. For leaves, this transition density seems to occur at higher densities.  \par

Interestingly, the leaves seem to have an overall sharper scaling behaviour compared to the branches. However, this separation cannot be seen in the Jeans length, where all structures show a consistent scaling of roughly $\lambda_{\rm B,T} \propto \rho^{-2/3}$ (Fig.~\ref{fig:magnetic_thermal_jeans_length}). This can be understood as follows: \par
The mass of a structure is dependent on the density and size, i.e.
\begin{equation}\label{eq:mass}
    M \propto \rho R^3.
\end{equation}
Combining Eq.~\ref{eq:mass} with Eq.~\ref{eq:magnetic_thermal_jeans_mass}, we obtain
\begin{equation}
    \frac{M}{M_{\rm B,T}} \propto R^3 \lambda_{\rm B,T}^{-3}.
\end{equation}
As the size of the leaf structures is more determined by the choice of $N_{\rm cells}$, they typically show very weak or no scaling between density and size, and we can therefore approximate $R \propto \rho^0$. For the leaves, this leads to $M/M_{\rm B,T} \propto \lambda_{\rm B,T}^{-3}$. As $\lambda_{\rm B,T} \propto \rho^{-2/3}$ approximately (Fig.~\ref{fig:magnetic_thermal_jeans_length}), this leads to $M/M_{\rm B,T} \propto \rho^2$. For the branches, we find a shallower slope. In \citet{ganguly_silcc-zoom_2022}, we find many branches to follow $M \propto R$, which would lead to $M/M_{\rm B,T} \propto \rho^{1/2}$, roughly consistent with the trend seen for the branches here. The relation of the scaling between $\lambda_{\rm B,T}$ and $\rho$ is in itself interesting and we discuss it in Appendix~\ref{sec:app_jeans_length_density}. \par
Overall, the Jeans analysis seems to show the emergence of potentially Jeans-unstable structures at slightly lower densities ($\sim 30$~cm$^{-3}$) compared to that found in the previous energetic analysis. This could reflect the fact that the Jeans analysis performed here does not include the kinetic energy, which is often larger compared to $E_{\rm B}$ and the thermal energy \citep[see Section 4 in][]{ganguly_silcc-zoom_2022}. Turbulent motions can act as an effective kinetic pressure term. Although the kinetic energy is often treated as an effective pressure in the literature \citep[see e.g.][]{chandrasekhar_gravitational_1951, bonazzola_jeans_1987, federrath_star_2012}, we show in \citet{ganguly_silcc-zoom_2022} that the volume and surface terms of the kinetic energy combine in a highly non-trivial manner, with structures often being confined or even compressed under ram pressure. This suggests that including a kinetic pressure in the Jeans analysis would be too simplistic and not meaningful.\par
Overall, most leaf fragments in the Jeans analysis have $M/M_{\rm B,T}<1$, suggesting that their fragmentation is unlikely to be primarily Jeans-like. However, above 10$^{-20}$~g~cm$^{-3}$, we begin to obtain Jeans unstable fragments which are mostly unresolved and will likely undergo further fragmentation, possibly ending up as the precursors of protostars.

\subsection{Delay introduced by magnetic fields}\label{sec:delay}
The fragmentation analysis performed in \ref{sec:fragmentation} seems to suggest that magnetic fields at least delay fragmentation in many cases. To estimate how much the evolution of the cloud is slowed down by the effect of magnetic fields, we define a delay timescale $\Delta t_{\rm B}$. 
Consider a structure of size $S$ which is compressed by an external flow with velocity $v$. In the absence of either gravity or magnetic fields, as well as neglecting internal thermal and kinetic pressure, 
the structure would be compressed on a crossing time,
\begin{equation}
 t_{\rm v} = S/v.   
\end{equation}
For simplicity, we estimate the size of a structure using the shortest axis of the equivalent ellipsoid, i.e. $S=2c$ (Section~\ref{sec:methods_structure_classification}). 
We approximate the sweep-up velocity $v$ to be equal to the bulk velocity of the structure, i.e. $v=|\mathbf{v}_0|$, where $\mathbf{v}_0$ is obtained from Eq.~\ref{eq:bulk_velocity}. So overall we have
\begin{equation}
 t_{\rm v} = 2c/|\mathbf{v}_0|.   
\end{equation}

Next, we consider an additional gravitational acceleration $a_{\rm g}$ assisting the sweeping up, where
\begin{equation}
    a_{\rm g} = -\frac{1}{V}\int_V \mathbf{g}\cdot\frac{\mathbf{r}-\mathbf{r}_0}{|\mathbf{r}-\mathbf{r}_0|} d^3r
\end{equation}
is the average acceleration towards the centre of mass, $\mathbf{r}_0$.
We can then estimate (to first order) the gravitationally assisted sweep-up timescale, $t_{\rm v,\ g}$, from 
\begin{equation}
    S = v t_{\rm v,\ g} + \frac{1}{2}a_{\rm g} t_{\rm v,\ g}^2.
\end{equation}
For non-gravitating structures, this reduces to $t_{\rm v}$. For non-zero gravitational field, taking the real root, we get
\begin{equation}\label{eq:t_v_g}
    t_{\rm v,g} = \frac{(-v + \sqrt{v^2 + 2Sa_{\rm g}})}{a_{\rm g}}.
\end{equation}

In the presence of magnetic fields, we can represent the combined acceleration by gravity and magnetic fields as $a_{\rm g,B}$, where
\begin{equation}
    a_{\rm g,B} = -\frac{1}{V}\int_V \left(\mathbf{g} - \frac{\nabla |\mathbf{B}|^2}{8\pi\rho}\right)\cdot\frac{\mathbf{r}-\mathbf{r}_0}{|\mathbf{r}-\mathbf{r}_0|} d^3r.
\end{equation}
We can then rewrite Eq.~\ref{eq:t_v_g} to estimate a combined timescale 
\begin{equation}
    t_{\rm v,g,B} = \frac{(-v + \sqrt{v^2 + 2Sa_{\rm g,B}})}{a_{\rm g,B}}.
\end{equation}
The time delay due to the presence of magnetic fields, $\Delta t_{\rm B}$, can then be estimated as
\begin{equation}\label{eq:delay_timescale}
    \Delta t_{\rm B} = t_{\rm v,g,B} - t_{\rm v,g}.
\end{equation}
\begin{figure}
    \centering
    \includegraphics[width=\columnwidth]{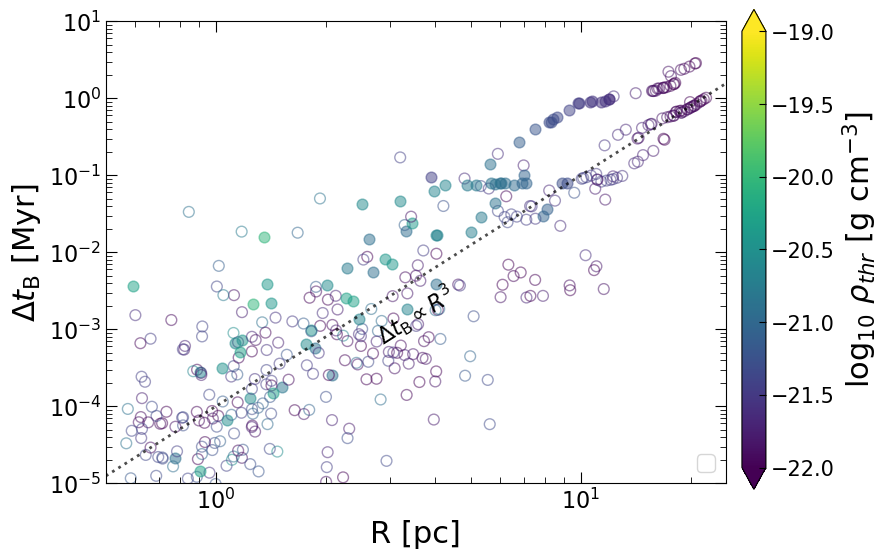}
    \caption{The estimated delay timescale, $\Delta t_{\rm B}$ (Eq.~\ref{eq:delay_timescale}), for various MHD cloud structures from the \textit{high-den} analysis at $t_{\rm evol}=2$~Myr. A power-law proportional to $R^3$ is plotted to show the rough scaling. }
    \label{fig:delay_timescale}
\end{figure}

In Fig.~\ref{fig:delay_timescale}, we plot $\Delta t_{\rm B}$ for various structures from the \textit{high-den} dendrogram analysis. We see that at the largest cloud scales at $t_{\rm evol}=2$~Myr, $\Delta t_{\rm B}$ is of the order of $\sim 1$~Myr, and then steadily decreases as a power-law roughly consistent with $\Delta t_{\rm B} \propto R^3$. This timescale of $\sim 1$~Myr seems to be consistent with the results of the fragmentation analysis in Section~\ref{sec:fragmentation}, where we found that the significant differences in the cloud fragmentation properties at $t_{\rm evol}=2$~Myr seem to have completely disappeared at $t_{\rm evol}=3.5$~Myr. The general power-law trend  We emphasise, however, that the calculation of $\Delta t_{\rm B}$ should only be considered a first-order approximation.
Note that $\Delta t_{\rm B}$ does not directly depend on the magnetic field strength but rather on its gradient. Hence, it is difficult to predict how $\Delta t_{\rm B}$ would scale with different strengths of the background field. This means that molecular clouds that form in a more magnetised medium do not necessarily form structures more slowly.   

\subsection{Densities at which magnetic fields become dynamically sub-dominant}
From the results presented in the previous sections, we can attempt to answer the question of at what densities magnetic fields become dynamically sub-dominant. From the density PDF of different clouds (Fig.~\ref{fig:density_pdf}), we find that the density distribution is significantly different in the presence of magnetic fields only below $\sim$~100~cm$^{-3}$. This is in accordance with previous simulations and observations \citep[][]{klessen_formation_2001, slyz_towards_2005, girichidis_evolution_2014, schneider_understanding_2015}, as well as the conclusions drawn by \citet{seifried_silcc-zoom_2020} using distributions of the three-dimensional true optical depth, $A_{\rm V,3D}$. From the energetic analysis (Fig.~\ref{fig:energetics}), we find that, magnitude-wise, gravity and kinetic energy supersede magnetic fields above a few $\sim 100$~cm$^{-3}$, consistent with the results of \citet{ibanez-mejia_gravity_2022}. Moreover, this density range is also in accordance with the results of \citet{seifried_parallel_2020}, who find that relative orientation of magnetic fields with respect to elongated filamentary structures changes at a few $\sim 100$~cm$^{-3}$ due to the occurrence of gravity-driven converging flows \citep{soler_what_2017}, suggesting energetic sub-dominance of magnetic fields at higher densities. Lastly, also the fragmentation analysis presented in this work (Fig.~\ref{fig:fragmentation}) shows differences in fragmentation patterns below a similar density regime of $\sim 100$~cm$^{-3}$. A Jeans fragmentation analysis yields roughly consistent limits as well.

In summary, for clouds born from an ISM with typical magnetic field strengths as in our Milky Way \citep{beck_magnetic_2013}, the density PDFs, the energetic analysis, the histogram of relative orientation technique applied by \citet{seifried_parallel_2020}, and the fragmentation analysis in this work - all seem to point to the fact that the magnetic field becomes sub-dominant above densities of
around $100 - 1000$~cm$^{-3}$. This overall trend is also fully consistent with the $B-\rho$ relation obtained by \citet{crutcher_magnetic_2010}, who conclude a transition density of $\sim 300$~cm$^{-3}$.

\section{Conclusions}\label{sec:conclusion}
We investigate the role magnetic fields play in determining the morphology, energetics, and fragmentation properties of young molecular clouds by analysing seven different simulated clouds (five with magnetic fields and two without) from the SILCC-Zoom simulations. These simulations are geared to study the evolution of the multi-phase interstellar medium in a supernova-driven, turbulent, stratified galactic disc environment. To identify forming structures, we use a dendrogram algorithm, and trace the statistical properties of the identified structures. We include a simple chemical network which allows us to follow the formation of H$_2$ as the cloud assembles and thereby distinguish between mostly atomic (H$_2$ mass fraction~<~50\%) and mostly molecular (H$_2$ mass fraction~>~50\%) structures. \par 
\begin{itemize}
    \item We observe that the MHD clouds are fluffier, meaning that they have more intermediate density gas between the number densities of roughly $1-100$~cm$^{-3}$, compared to their hydrodynamic counterparts. In the hydrodynamic clouds, the lack of magnetic fields results in the denser structures being surrounded by a comparatively more rarefied envelope. \par

    \item In terms of morphology, we find that almost all clouds are sheet-like, which is consistent with recent observations of sheet-like envelopes around denser filamentary cloud structures \citep[][]{kalberla_cold_2016-1, arzoumanian_molecular_2018, rezaei_kh_three-dimensional_2022, tritsis_musca_2022, pineda_bubbles_2022, clarke_herschel_2023}. In our case, the MCs form due to compressions caused by expanding supernova shells, consistent with the bubble-driven MC formation scenario \citep[][]{koyama_molecular_2000,inoue_two-fluid_2009,inutsuka_formation_2015}. \par 
    \item We find that spheroidal structures within the clouds are rare on all spatial scales, with $\sim 90$\% of the structures being elongated. We further see that the runs with magnetic fields have a roughly comparable fraction of filaments and sheets, whereas the hydrodynamic runs overall produce more sheet-like structures compared to filaments. \par
    \item Energetically, magnetic fields in our simulations are important for less dense (up to $\sim$1000~cm$^{-3}$) and mostly, but not exclusively, atomic structures. The dynamics for denser and potentially star-forming structures is dominated by the interplay of turbulence and gravity. This density threshold, above which the magnetic fields seems to become sub-dominant, is supported by the previous works of \citet{seifried_parallel_2020}, \citet{ibanez-mejia_gravity_2022} and is consistent with the observed transition in the $B-\rho$ relation \citep[][]{crutcher_magnetic_2010}. 
    \item By investigating the magnetic surface energy term, we find that for most structures it acts in a confining manner, and, for some low-density structures, it even leads to overall magnetic confinement.
\par
    \item By studying the numbers and masses of cloud fragments that form, we find that at densities below roughly $\sim 100$~cm$^{-3}$, the presence of magnetic fields helps to create more massive fragments, but generally does not result in an increase in the number of such structures. A stability analysis suggests that in the resolved range, leaf fragments are mostly Jeans stable and the fragmentation is not primarily governed by magnetic Jeans instabilities. Instead of significantly altering the nature of fragmentation, magnetic fields seem to rather slow down the fragmentation process. Using a simple order-of-magnitude estimate, we find that this delay timescale is $\sim 1$~Myr. 
\par
\end{itemize}
Overall, using density PDFs, and an energetic as well as a fragmentation analysis, we find a scenario where magnetic fields significantly affect the flows and fragmentation in the lower density gas (below $\sim 100$~cm$^{-3}$), channelling flows and thereby affecting both, the morphology of the forming structures as well as the formation timescale of the dense gas. Once the dense structures (typically above $\sim 1000$~cm$^{-3}$) form, however, the further evolution and fragmentation of the dense gas seems to be mostly unaffected by the magnetic field. 

\section*{Acknowledgements}
We would like to thank the referee, Prof.~Dr. Robi Banerjee, for their helpful comments, suggestions, and overall discussion, which have increased the quality of the paper. SG, SW, DS and MW would like to acknowledge the support of Bonn-Cologne Graduate School (BCGS), which is funded through the German Excellence Initiative, as well as the DFG for funding through SFB 956 'Conditions and Impact of Star Formation' (subprojects C5 and C6). SDC is supported by the Ministry of Science and Technology (MoST) in Taiwan through grant MoST 108-2112-M-001-004-MY2. This research made use of \texttt{astrodendro}, a Python package to compute dendrograms of Astronomical data (http://www.dendrograms.org/); as well as \texttt{yt}, an open-source, permissively-licensed python package for analyzing and visualizing volumetric data (\href{https://yt-project.org/}{https://yt-project.org/}). The 3D renderings in Fig.~\ref{fig:3d_projection} were computed using \texttt{paraview}. The FLASH code used in this work was partly developed by the Flash Center for Computational Science at the University of Rochester.

\section*{Data Availability}
The data underlying this article can be shared for selected scientific purposes after request to the corresponding author.



\bibliographystyle{mnras}
\bibliography{template} 




\appendix

\section{Basic information of clouds}\label{sec:app_a}
\begin{figure*}
\includegraphics[width=0.99\textwidth]{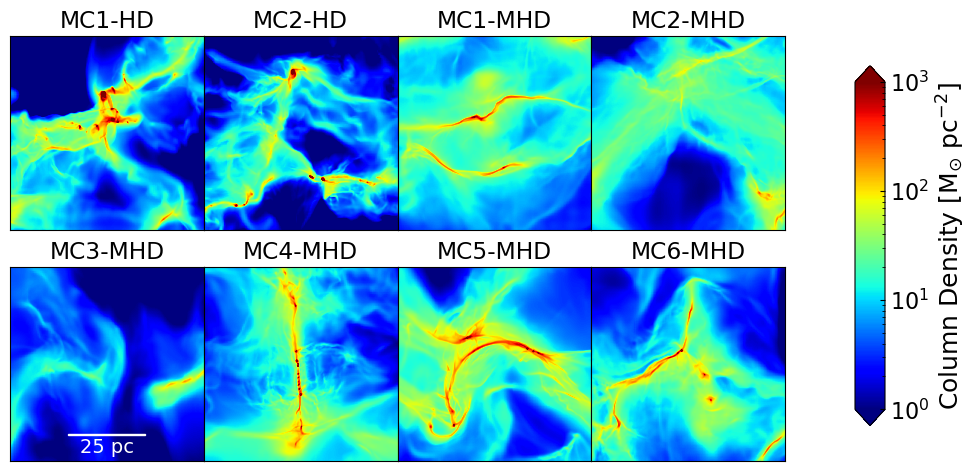}
    \caption{Column density projection along the x axis for different molecular clouds at t$_{\rm evol}=3.5$~Myr. The MHD clouds have typically more diffuse emission. Note that we have excluded MC3-MHD from further analysis due to its low molecular content (see Fig. \ref{fig:h2_mass}).}
    \label{fig:projection_plot}
\end{figure*}

\begin{figure*}
    \centering
    \includegraphics[trim=5 0 5 0, clip,width=0.33\textwidth]{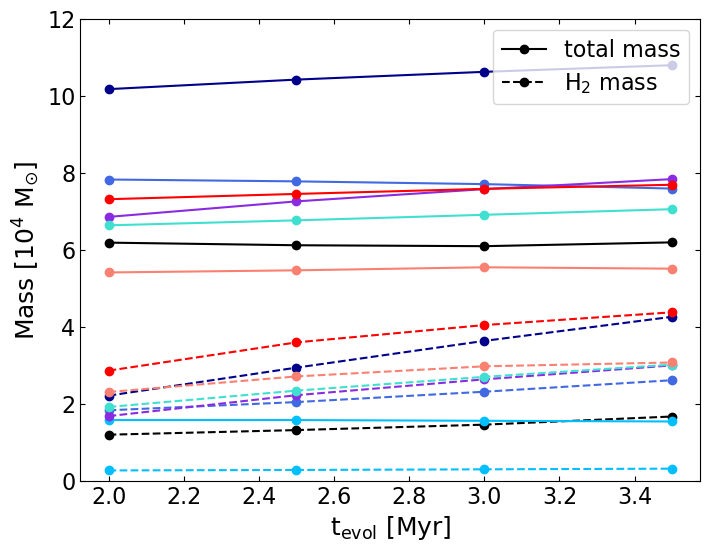}
    \includegraphics[trim=5 0 5 0, clip,width=0.33\textwidth]{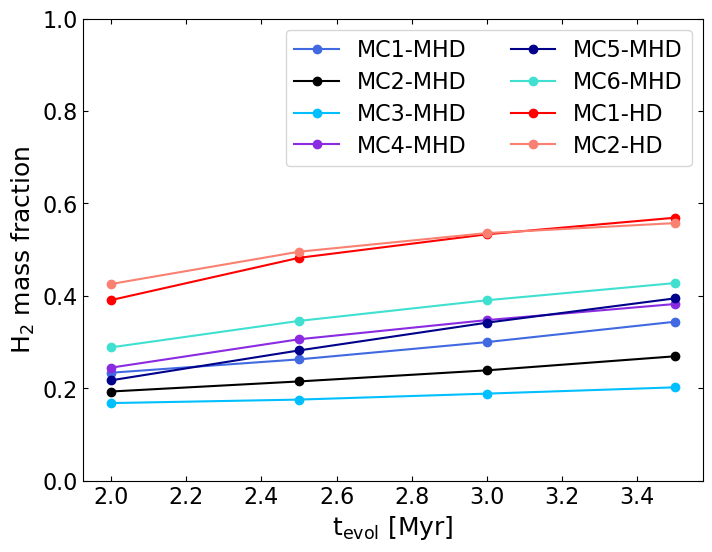}
    \includegraphics[trim=5 0 5 0, clip,width=0.33\textwidth]{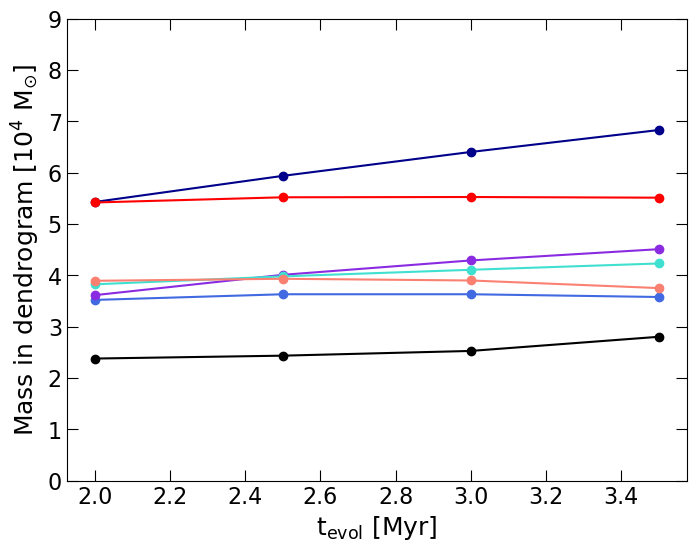}
    \caption{Left: Time evolution of total mass and total H$_2$ mass in the different molecular clouds, both HD and MHD, from $t_{\rm evol}=2$ to 3.5~Myr. The solid lines represent the total mass, and the dashed lines represent the H$_2$ mass. Middle: H$_2$ mass fraction for the same clouds, both HD and MHD. The two HD clouds are plotted in reddish lines. Apart from MC3-MHD, which we discard due to its low molecular gas mass, the other MHD and HD and clouds have comparable masses. The two HD clouds, however, have a much higher H$_2$ mass fraction. Right: The mass in dendrogram for each cloud. Note that the discarded MC3-MHD is missing as we did not perform a dendrogram on it. The dendrogram mass has similar trends to the total mass.}
    \label{fig:h2_mass}
\end{figure*}

We present here some basic properties of the different analysed molecular clouds. Fig.~\ref{fig:projection_plot} plots the column density projections of all different clouds, both HD and MHD. Fig.~\ref{fig:h2_mass} plots the total and H$_2$ mass of the different MCs in the left panel, and the H$_2$ mass fraction in the right panel. We see that there is no difference in the overall mass of the clouds depending on the magnetic field, but that the H$_2$ mass fraction in the HD clouds is higher. The cloud MC3-MHD (cyan line), which has been excluded from this analysis, has the lowest total H$_2$ mass, as well as the lowest H$_2$ mass fraction.

\section{Alternative PDF views}\label{sec:app_pdf}
\begin{figure}
    \centering
    \includegraphics[width=0.49\textwidth]{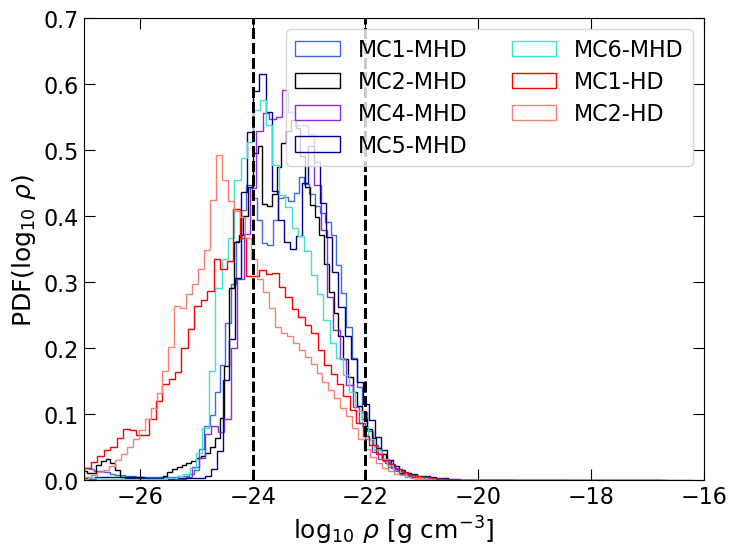}
    \caption{Density PDF with a linear y-axis for all HD and MHD clouds at t$_{\rm evol}$=3.5~Myr.}
    \label{fig:pdf_and_av}
\end{figure}

\begin{figure}
    \centering
    \includegraphics[width=0.49\textwidth]{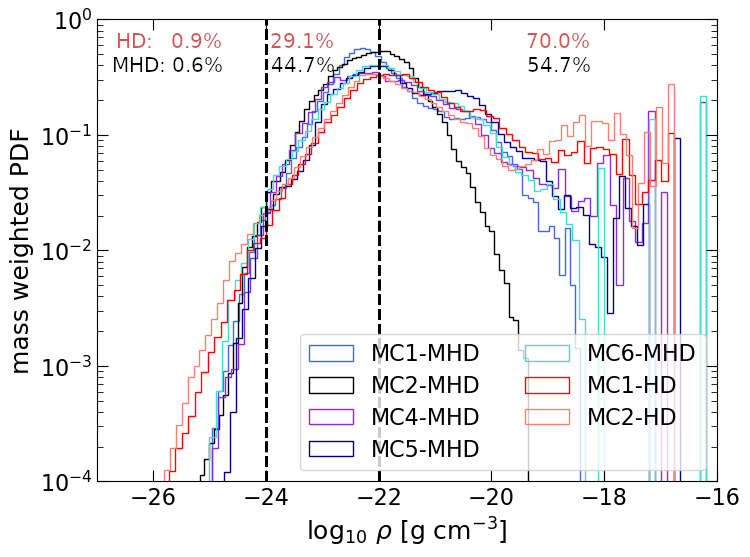}
    \caption{Mass-weighted PDF for all clouds, both HD and MHD, at t$_{\rm evol}=3.5$~Myr. The average mass percentage in the different regimes (shown by the vertical dotted bars), for both HD and MHD clouds, is shown as text. The mass contained at $\rho < 10^{-24}$~g~cm$^{-3}$ is $<1\%$ for all clouds. The mass difference in the intermediate regime, even at 3.5 Myr, is clearly seen.}
    \label{fig:mass_weighted_pdf}
\end{figure}

We present additional views of density PDFs (both with a linear scale and mass-weighted) in Figs.~\ref{fig:pdf_and_av} and \ref{fig:mass_weighted_pdf}  as a complementary addition to Fig.~\ref{fig:density_pdf}.

\section{Distribution of visual extinction for magnetized and non-magnetized clouds}\label{sec:app_av}
\begin{figure}
    \centering
    \includegraphics[width=0.49\textwidth]{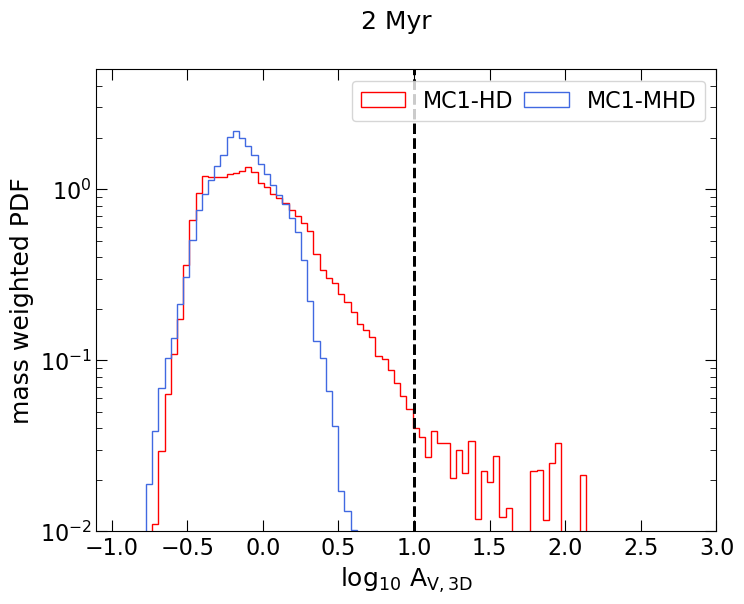}
    \includegraphics[width=0.49\textwidth]{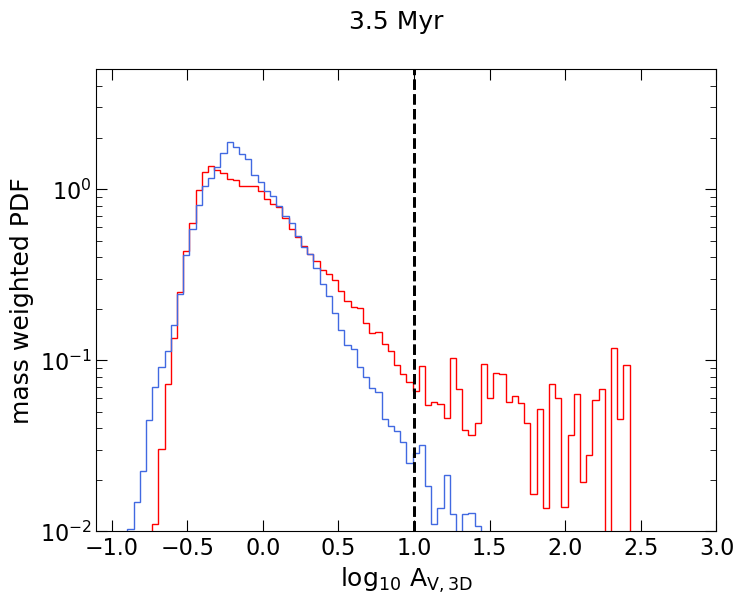}
    \caption{Mass weighted $\mathrm{A_{v,3D}}$ PDF for different HD and MHD clouds at $t_{\rm evol}$=3.5~Myr. MC2-MHD stands out as having much less shielded gas mass compared to the other clouds. The other HD and MHD clouds have similar behaviour.}
    \label{fig:av_pdf}
\end{figure}

The gas mass distribution at different $\mathrm{A_{V,3D}}$ values (Eq.~\ref{eq:av_3d}) for one example HD and MHD cloud of comparable mass is presented in Fig.~\ref{fig:av_pdf}, for 2~Myr (top) and 3.5~Myr (bottom). The vertical dashed line represents $\mathrm{A_{V,3D}}=1$. The HD cloud has consistently higher mass at high $\mathrm{A_{V,3D}}$ values.

\section{Alternative view of the magnetic field - density relation}
As a companion view to Fig.\ref{fig:bn_relation}, we show here the same relation between the magnetic field and density, but this time using the average density, $\rho_{\rm avg}$ instead of $\rho_{\rm thr}$. Since for any given structure, $\rho_{\rm avg} \geq \rho_{\rm thr}$, this results in a shallower fit at the high density end using $\rho_{\rm avg}$.
\begin{figure}
    \centering
    \includegraphics[width=\columnwidth]{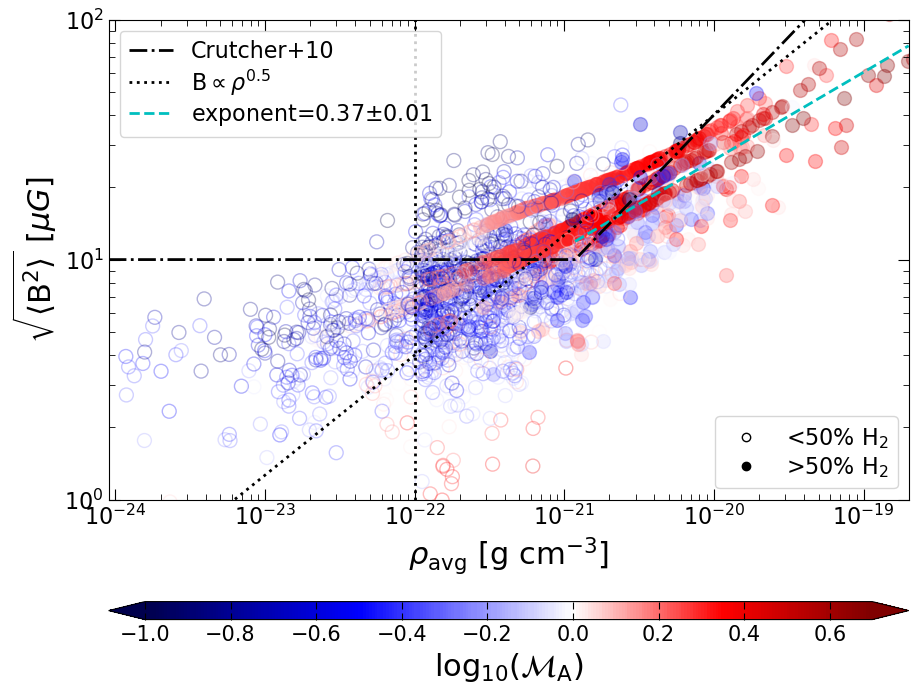}
    \caption{Similar to Fig.~\ref{fig:bn_relation}, but using $\rho_{\rm avg}$ instead of $\rho_{\rm thr}$. This creates a shallower slope at particularly the high density end, as $\rho_{\rm avg} \geq \rho_{\rm thr}$. However, the overall statistical trend is similar.}
    \label{fig:bn_relation_avg_den}
\end{figure}

\section{Supplement to the magnetic Jeans analysis}\label{sec:app_magnetic_jeans}
\begin{figure}
    \includegraphics[width=\columnwidth]{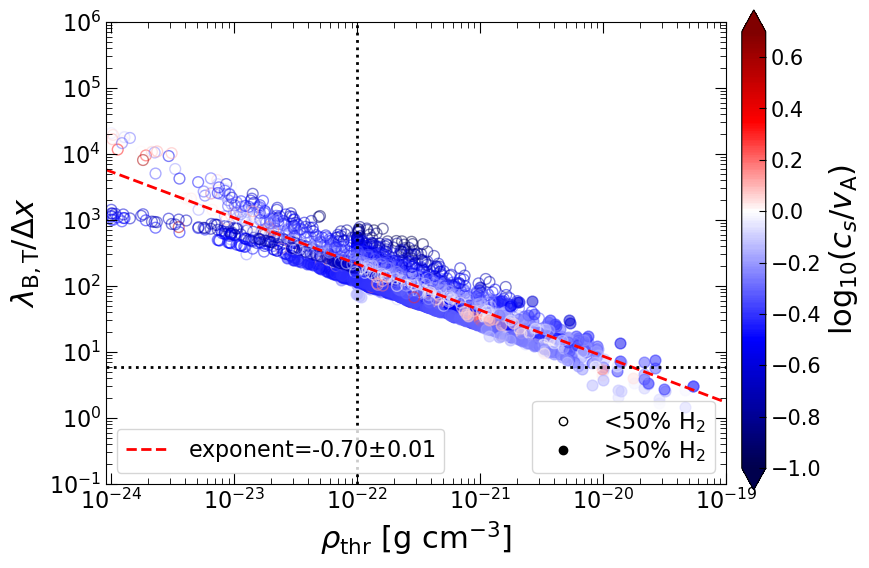}
    \caption{The ratio of the magneto-thermal Jeans length, $\lambda_{\rm B,T}$ to the maximum resolution $\Delta x$ ($\sim 0.125$~pc for $\rho_{\rm thr}>10^{-22}$~g~cm$^{-3}$ and $\sim 0.25$~pc for $\rho_{\rm thr}<10^{-22}$~g~cm$^{-3}$, see also Table~\ref{tab:dendrogram_type}), as a function of $\rho_{\rm thr}$ for all MHD sub-structures at 3.5~Myr. The horizontal dotted line denotes the resolution limit for the present dendrogram analysis ($2N_{\rm cells}^{1/3}$, with $N_{\rm cells}=100$). The red dashed line denotes the best-fit exponent for a linear least-squares fit on the logarithm of the data. Structures above $\rho_{\rm thr} \approx 10^{-20}$~g~cm$^{-3}$ seem to be not well resolved enough to be conclusive regarding the fragmentation analysis.}
    \label{fig:magnetic_thermal_jeans_length}
\end{figure}

The Jeans mass analysis is only conclusive provided the Jeans length is resolved. In Fig.~\ref{fig:magnetic_thermal_jeans_length}, we plot the ratio of the magneto-thermal Jeans length, $\lambda_{\rm B,T}$, to the maximum resolution, $\Delta x$ ($\sim 0.125$~pc for $\rho_{\rm thr}>10^{-22}$~g~cm$^{-3}$ and $\sim 0.25$~pc for $\rho_{\rm thr}<10^{-22}$~g~cm$^{-3}$, see also Table~\ref{tab:dendrogram_type}), as a function of $\rho_{\rm thr}$ for all sub-structures at 3.5~Myr. The colour-bar, similar to Fig.~\ref{fig:magnetic_thermal_jeans}, denotes $c_s/v_{\rm A}$. The horizontal dotted line denotes $(2N_{\rm cells})^{1/3}$. As $N_{\rm cells}$ denotes the minimum number of cells required in the dendrogram analysis for any structure, $2N_{\rm cells}$ is the minimum number of cells a structure must have in order to fragment. Therefore, $(2N_{\rm cells})^{1/3}$ represents the minimum number of cells required in one direction by which the Jeans length should be resolved. We find that this seems to not be the case only for some structures with $\rho_{\rm thr}\gtrsim 10^{-20}$~g~cm$^{-3}$. When we fit $\lambda_{\rm B,T}$ against $R$ using a linear least-squares fit on the logarithm of the data, we obtain an exponent of $-0.70 \pm 0.01$, roughly consistent with an exponent of $-2/3$. \par

\begin{figure}
    \includegraphics[width=\columnwidth]{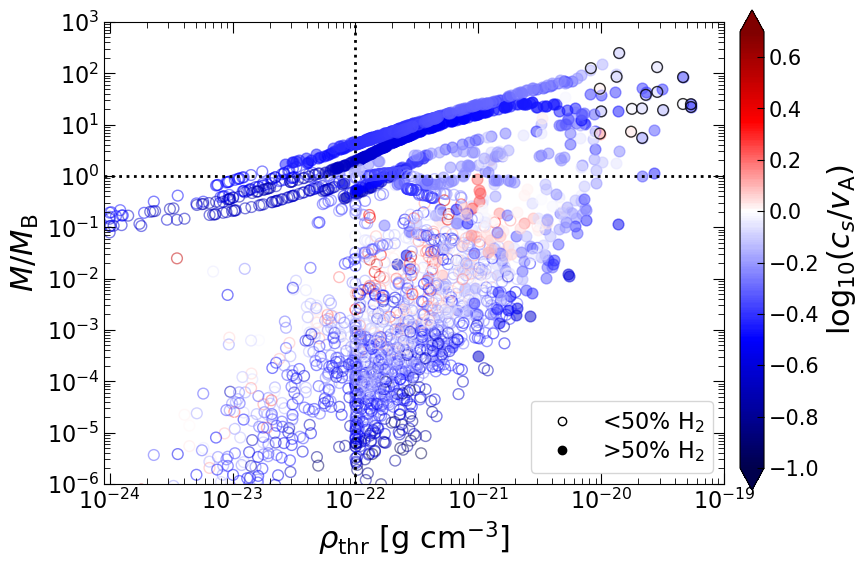}
    \includegraphics[width=\columnwidth]{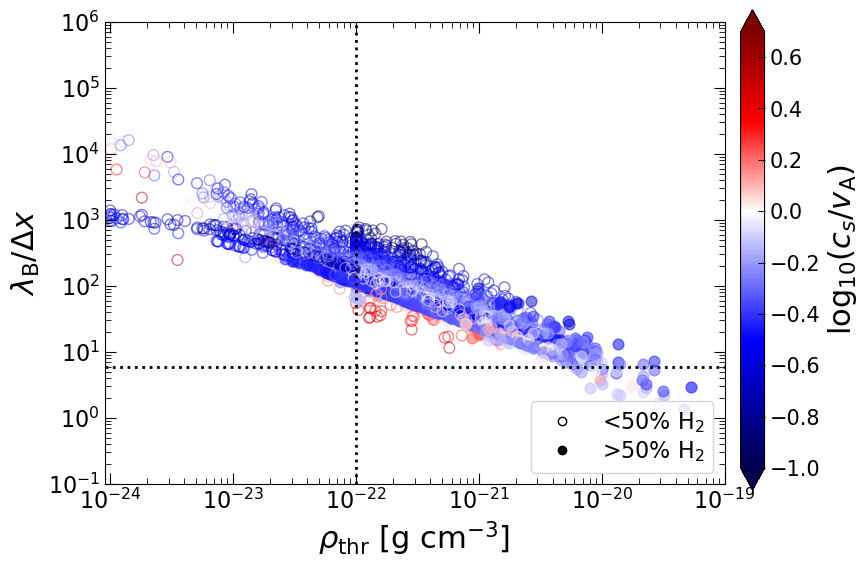}    
    \caption{Top: same as the combined panels of Fig.~\ref{fig:magnetic_thermal_jeans}, but for a purely magnetic Jeans mass. Bottom: same as Fig.~\ref{fig:magnetic_thermal_jeans_length}, but for a purely magnetic Jeans length.}
    \label{fig:magnetic_jeans}
\end{figure}

\begin{figure}
    \includegraphics[width=\columnwidth]{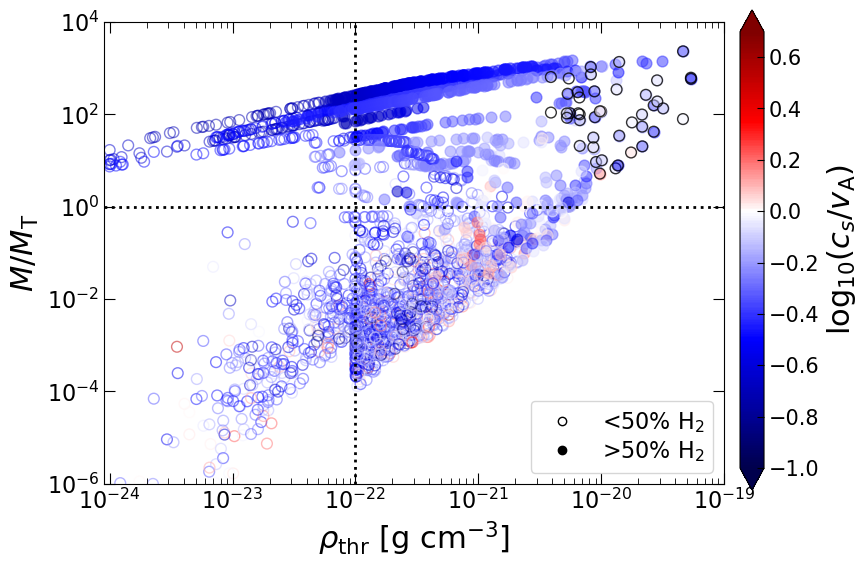}
    \includegraphics[width=\columnwidth]{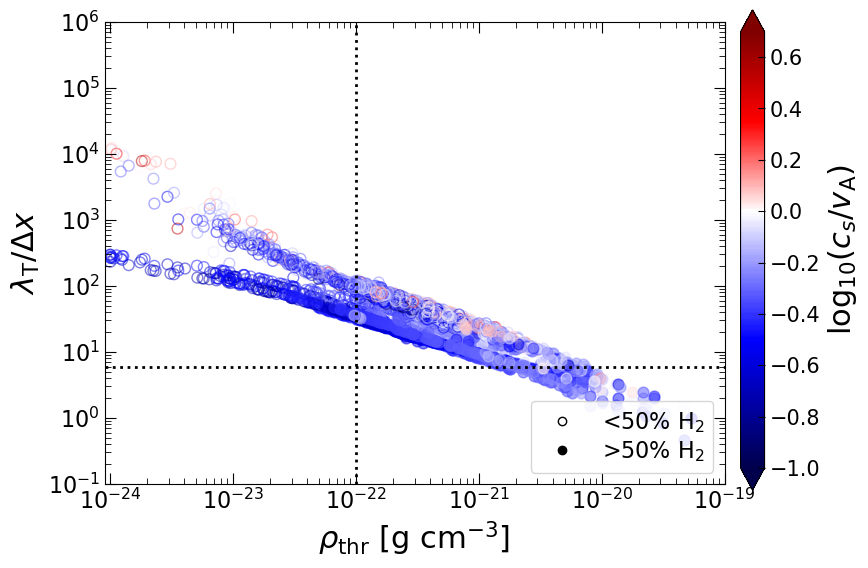}    
    \caption{Top: same as Fig.~\ref{fig:magnetic_jeans}, but for a purely thermal Jeans mass. Bottom: same as Fig.~\ref{fig:magnetic_thermal_jeans_length}, but for a purely thermal Jeans length. Note that the $y$-range in the top panel is different to the previous similar plots.}
    \label{fig:thermal_jeans}
\end{figure}

The analysis performed in Fig.~\ref{fig:magnetic_thermal_jeans} considers the combined contribution of magnetic and thermal perturbations. It might be interesting to note their relative contributions. For this purpose, we show a purely magnetic Jeans analysis in Fig.~\ref{fig:magnetic_jeans}. Comparing $M/M_{\rm B}$ to $M/M_{\rm B,T}$ (Fig.~\ref{fig:magnetic_thermal_jeans}), we find little to no difference, suggesting that the magnetic contribution is in this density range more important than the thermal contribution. This can also be seen in the fact that most of the points have larger $v_{\rm A}$ compared to $c_s$ (bluish in the colour-bar). For completeness, we explicitly include the thermal Jeans mass and length plot in Fig.~\ref{fig:thermal_jeans}. 

\section{The scaling-relation between Jeans length and density}\label{sec:app_jeans_length_density}
The Jeans length, $\lambda$, depends on the characteristic wave speed, $c$ ($c_s$, $v_{\rm A}$, or a combination of the two), and the density, i.e.
\begin{equation}
    \lambda \propto c \rho^{-1/2}.
\end{equation}
In our case, we are dominated by the magnetic over kinetic pressure, i.e. $\lambda_{\rm B,T} \approx \lambda_{\rm B}$. The Alfv\'en wave speed scales as
\begin{equation}
    v_{\rm A} \propto \frac{B}{\rho^{1/2}}.
\end{equation}
This leads to
\begin{equation}
    \lambda_{\rm B,T} \propto \frac{B}{\rho}.
\end{equation}
For a scaling of $B \propto \rho^{1/2}$, this leads to $\lambda_{\rm B,T} \propto \rho^{-1}$. A scaling of $B \propto \rho^{1/3}$ leads to $\lambda_{\rm B,T} \propto \rho^{-2/3}$. The fitted value seems to be somewhere in-between, closer to $\rho^{-2/3}$, and is also roughly consistent with the overall $B-\rho$ scaling in Fig.~\ref{fig:bn_relation} and Fig.~\ref{fig:bn_relation_avg_den}.

\bsp	
\label{lastpage}
\end{document}